\documentclass[a4paper,11pt]{article}
\pdfoutput=1 

\usepackage{jcappub} 

\usepackage[T1]{fontenc} 
\usepackage{aasmacros}

\usepackage{amsmath}
\usepackage{bm}
\usepackage{tabularx}
\usepackage{mleftright}

\usepackage{booktabs}

\usepackage{caption}
\usepackage{subcaption}

\usepackage{xcolor}
\usepackage{soul} 

\title{\boldmath 
Polarization angle requirements for CMB B-mode experiments. Application to the LiteBIRD satellite}

\author[1]{P.\,Vielva,}
\author[1]{E.\,Martínez-González,}
\author[1]{F.\,J.\,Casas,}
\author[2]{T.\,Matsumura,}
\author[3]{S.\,Henrot-Versillé,}
\author[4,2]{E.\,Komatsu,}
\author[5]{J.\,Aumont,}
\author[6]{R.\,Aurlien,}
\author[7,8,9]{C.\,Baccigalupi,}
\author[5]{A.\,J.\,Banday,}
\author[1]{R.\,B.\,Barreiro,}
\author[10,11,12]{N.\,Bartolo,}
\author[13]{E.\,Calabrese,}
\author[14,15,16]{K.\,Cheung,}
\author[17,18]{F.\,Columbro,}
\author[17,18]{A.\,Coppolecchia,}
\author[17,18]{P.\,de\,Bernardis,}
\author[19]{T.\,de\,Haan,}
\author[1,20]{E.\,de\,la\,Hoz,}
\author[17,18]{M.\,De\,Petris,}
\author[21]{S.\,Della\,Torre,}
\author[1,20]{P.\,Diego-Palazuelos,}
\author[6]{H.\,K.\,Eriksen,}
\author[22]{J.\,Errard,}
\author[23,24]{F.\,Finelli,}
\author[25,26]{C.\,Franceschet,}
\author[6]{U.\,Fuskeland,}
\author[6]{M.\,Galloway,}
\author[22]{K.\,Ganga,}
\author[27,21]{M.\,Gervasi,}
\author[28,29]{R.\,T.\,Génova-Santos,}
\author[30,2]{T.\,Ghigna,}
\author[6]{E.\,Gjerløw,}
\author[23,24]{A.\,Gruppuso,}
\author[31,19,32,2,33]{M.\,Hazumi,}
\author[1]{D.\,Herranz,}
\author[34]{E.\,Hivon,}
\author[19]{K.\,Kohri,}
\author[17,18]{L.\,Lamagna,}
\author[22]{C.\,Leloup,}
\author[35]{J.\,Macias-Perez,}
\author[17,18]{S.\,Masi,}
\author[32]{F.\,T.\,Matsuda,}
\author[23]{G.\,Morgante,}
\author[36,32]{R.\,Nakano,}
\author[27,21]{F.\,Nati,}
\author[37,38]{P.\,Natoli,}
\author[39]{S.\,Nerval,}
\author[32]{K.\,Odagiri,}
\author[32]{S.\,Oguri,}
\author[37,38,40]{L.\,Pagano,}
\author[17,18]{A.\,Paiella,}
\author[23,24]{D.\,Paoletti,}
\author[17,18]{F.\,Piacentini,}
\author[41]{G.\,Polenta,}
\author[42,16]{G.\,Puglisi,}
\author[1,43]{M.\,Remazeilles,}
\author[44,40,45]{A.\,Ritacco,}
\author[28,29]{J.\,A.\,Rubino-Martin,}
\author[46]{D.\,Scott,}
\author[32,36,19]{Y.\,Sekimoto,}
\author[47]{M.\,Shiraishi,}
\author[48]{G.\,Signorelli,}
\author[36,32]{H.\,Takakura,}
\author[48,49]{A.\,Tartari,}
\author[50,51]{K.\,L.\,Thompson,}
\author[3]{M.\,Tristram,}
\author[5]{L.\,Vacher,}
\author[42,52]{N.\,Vittorio,}
\author[6]{I.\,K.\,Wehus,}
\author[27,21]{and M.\,Zannoni}
\author[ ]{\\LiteBIRD Collaboration.}
\affiliation[1]{Instituto de Fisica de Cantabria (IFCA, CSIC-UC), Avenida los Castros SN, 39005, Santander, Spain}
\affiliation[2]{Kavli Institute for the Physics and Mathematics of the Universe (Kavli IPMU, WPI), UTIAS, The University of Tokyo, Kashiwa, Chiba 277-8583, Japan}
\affiliation[3]{Universit\'e Paris-Saclay, CNRS/IN2P3, IJCLab, 91405 Orsay, France}
\affiliation[4]{Max Planck Institute for Astrophysics, Karl-Schwarzschild-Str. 1, D-85748 Garching, Germany}
\affiliation[5]{IRAP, Universit$\acute{\rm e}$ de Toulouse, CNRS, CNES, UPS, (Toulouse), France}
\affiliation[6]{Institute of Theoretical Astrophysics, University of Oslo, Blindern, Oslo, Norway}
\affiliation[7]{International School for Advanced Studies (SISSA), Via Bonomea 265, 34136, Trieste, Italy}
\affiliation[8]{INFN Sezione di Trieste, via Valerio 2, 34127 Trieste, Italy}
\affiliation[9]{IFPU, Via Beirut, 2, 34151 Grignano, Trieste, Italy}
\affiliation[10]{Dipartimento di Fisica e Astronomia “G. Galilei”, Universita` degli Studi di Padova, via Marzolo 8, I-35131 Padova, Italy}
\affiliation[11]{INFN Sezione di Padova, via Marzolo 8, I-35131, Padova, Italy}
\affiliation[12]{INAF, Osservatorio Astronomico di Padova, Vicolo dell’Osservatorio 5, I-35122, Padova, Italy}
\affiliation[13]{Cardiff University, School of Physics and Astronomy, Cardiff CF10 3XQ, UK}
\affiliation[14]{University of California, Berkeley, Department of Physics, Berkeley, CA 94720, USA}
\affiliation[15]{University of California, Berkeley, Space Sciences Laboratory,  Berkeley, CA 94720, USA}
\affiliation[16]{Lawrence Berkeley National Laboratory (LBNL), Computational Cosmology Center, Berkeley, CA 94720, USA}
\affiliation[17]{Dipartimento di Fisica, Universit\`{a} La Sapienza, P. le A. Moro 2, Roma, Italy}
\affiliation[18]{INFN Sezione di Roma, P.le A. Moro 2, 00185 Roma, Italy}
\affiliation[19]{Institute of Particle and Nuclear Studies (IPNS), High Energy Accelerator Research Organization (KEK), Tsukuba, Ibaraki 305-0801, Japan}
\affiliation[20]{Dpto. de Física Moderna, Universidad de Cantabria, Avda. los Castros s/n, E-39005 Santander, Spain}
\affiliation[21]{INFN Sezione Milano Bicocca, Piazza della Scienza, 3, 20126 Milano, Italy}
\affiliation[22]{Université de Paris, CNRS, Astroparticule et Cosmologie, F-75013 Paris, France}
\affiliation[23]{INAF - OAS Bologna, via Piero Gobetti, 93/3, 40129 Bologna (Italy)}
\affiliation[24]{INFN Sezione di Bologna, Viale C. Berti Pichat, 6/2 – 40127 Bologna Italy}
\affiliation[25]{Dipartimento di Fisica, Universita' degli Studi di Milano, Via Celoria 16 - 20133, Milano, Italy}
\affiliation[26]{INFN Sezione di Milano, Via Celoria 16 - 20133, Milano, Italy}
\affiliation[27]{University of Milano Bicocca, Physics Department, p.zza della Scienza, 3, 20126 Milan Italy}
\affiliation[28]{Instituto de Astrofísica de Canarias, E-38200 La Laguna, Tenerife, Canary Islands, Spain}
\affiliation[29]{Departamento de Astrofísica, Universidad de La Laguna (ULL), E-38206, La Laguna, Tenerife, Spain}
\affiliation[30]{Department of Physics, University of Oxford, Denys Wilkinson Building, Keble Road, Oxford OX1 3RH, United Kingdom}
\affiliation[31]{International Center for Quantum-field Measurement Systems for Studies of the Universe and Particles (QUP), High Energy Accelerator Research Organization (KEK), Tsukuba, Ibaraki 305-0801, Japan}
\affiliation[32]{Japan Aerospace Exploration Agency (JAXA), Institute of Space and Astronautical Science (ISAS), Sagamihara, Kanagawa 252-5210, Japan}
\affiliation[33]{The Graduate University for Advanced Studies (SOKENDAI), Miura District, Kanagawa 240-0115, Hayama, Japan}
\affiliation[34]{Institut d'Astrophysique de Paris, CNRS/Sorbonne Universit$\acute{\rm e}$, Paris France}
\affiliation[35]{Univ. Grenoble Alpes, CNRS, LPSC-IN2P3, 53, avenue des Martyrs, 38000 Grenoble, France}
\affiliation[36]{The University of Tokyo, Department of Astronomy, Tokyo 113-0033, Japan}
\affiliation[37]{Dipartimento di Fisica e Scienze della Terra, Universit\`a di Ferrara, Via Saragat 1, 44122 Ferrara, Italy}
\affiliation[38]{INFN Sezione di Ferrara, Via Saragat 1, 44122 Ferrara, Italy}
\affiliation[39]{David A. Dunlap Department of Astronomy and Astrophysics, 50 St. George Street, Toronto ON M5S3H4}
\affiliation[40]{Université Paris-Saclay, CNRS, Institut d’Astrophysique Spatiale, 91405, Orsay, France.}
\affiliation[41]{Space Science Data Center, Italian Space Agency, via del Politecnico, 00133, Roma, Italy}
\affiliation[42]{Dipartimento di Fisica, Universit\`{a} di Roma Tor Vergata, Via della Ricerca Scientifica, 1, 00133, Roma, Italy}
\affiliation[43]{Jodrell Bank Centre for Astrophysics, Alan Turing Building, Department of Physics and Astronomy, School of Natural Sciences, The University of Manchester, Oxford Road, Manchester M13 9PL, UK}
\affiliation[44]{INAF-Osservatorio Astronomico di Cagliari, Via della Scienza 5, 09047 Selargius, IT}
\affiliation[45]{Laboratoire de Physique de l’$\acute{\rm E}$cole Normale Sup$\acute{\rm e}$rieure, ENS, Universit$\acute{\rm e}$ PSL, CNRS, Sorbonne Universit$\acute{\rm e}$, Universit$\acute{\rm e}$ de Paris, 75005 Paris, France}
\affiliation[46]{Department of Physics and Astronomy, University of British Columbia, 6224 Agricultural Road, Vancouver BC, V6T1Z1, Canada}
\affiliation[47]{National Institute of Technology, Kagawa College}
\affiliation[48]{INFN Sezione di Pisa, Largo Bruno Pontecorvo 3, 56127 Pisa (Italy)}
\affiliation[49]{Dipartimento di Fisica, Università di Pisa, Largo B. Pontecorvo 3, 56127 Pisa}
\affiliation[50]{SLAC National Accelerator Laboratory, Kavli Institute for Particle Astrophysics and Cosmology (KIPAC),  Menlo Park, CA 94025, USA}
\affiliation[51]{Stanford University, Department of Physics,  CA 94305-4060, USA}
\affiliation[52]{INFN Sezione di Roma2, Universit\`{a} di Roma Tor Vergata, via della Ricerca Scientifica, 1, 00133 Roma, Italy}

\emailAdd{vielva@ifca.unican.es}

\abstract{
A methodology to provide the polarization angle requirements for different sets of detectors, at a given frequency of a CMB polarization experiment, is presented. The uncertainties in the polarization angle of each detector set are related to a given bias on the tensor-to-scalar ratio $r$ parameter. The approach is grounded in using a linear combination of the detector sets to obtain the CMB polarization signal. In addition, assuming that the uncertainties on the polarization angle are in the small angle limit (lower than a few degrees), it is possible to derive analytic expressions to establish the requirements. The methodology also accounts for possible correlations among detectors, that may originate from the optics, wafers, etc. The approach is applied to the LiteBIRD space mission. We show that, for the most restrictive case (i.e., full correlation of the polarization angle systematics among detector sets), the requirements on the polarization angle uncertainties are of around 1 arcmin at the most sensitive frequency bands (i.e., $\approx 150$ GHz) and of few tens of arcmin at the lowest (i.e., $\approx 40$ GHz) and highest (i.e., $\approx 400$ GHz) observational bands. Conversely, for the least restrictive case (i.e., no correlation of the polarization angle systematics among detector sets), the requirements are $\approx 5$ times 
less restrictive than for the previous scenario. At the global and the telescope levels, polarization angle knowledge  of a few arcmins is sufficient for correlated global systematic errors and can be relaxed by a factor of two for fully uncorrelated errors in detector polarization angle. The reported uncertainty levels are needed in order to have the  bias on $r$ due to systematics below  the limit established by the LiteBIRD collaboration.}

\begin{document}
\maketitle
\flushbottom

\section{Introduction}
\label{sec:intro}
The quest for the detection of the primordial gravitational waves (PGW) originated from the inflationary period of the early universe is one of the major goals in cosmology \cite{Gri74,Star79}. It is widely recognized that, given the amplitude and shape of the spectrum predicted in the standard models of inflation, the best strategy to measure these tiny disturbances in the spacetime curvature is in an indirect way through their imprint in the polarization B-mode of the Cosmic Microwave Background (CMB) \cite{Kam15}. Observational evidence of the existence of the PGW would represent a milestone in our knowledge of the physics of the  universe, providing a major, but still missing evidence that the universe went through an inflationary episode in the early universe \cite{Ly96,Ly98}.

Currently the most stringent limit on the amplitude of the PGW, represented by the tensor-to-scalar parameter $r$, is $r<0.032$ (95\% CL) obtained combining Planck and BICEP2/Keck Array data 
\cite{Trist21imp} (see also \cite{BicepKeck2021,Tristram2021,Planck2018,BicepKeck2018}). 

There is currently a large number of experiments from ground, stratospheric balloons or space-based, such as the BICEP array  \cite{Hui2018}, the Simons Observatory \cite{Ade2019}, the CMB-S4 experiment \cite{Abazajian2016} or the JAXA LiteBIRD satellite  \cite{Hazumi2019}, that are being operated or in development and with sensitivity to detect PGW with amplitudes smaller than the current limit. In particular, the LiteBIRD satellite and the CMB-S4 will probe the existence of the PGW background at sensitivity levels corresponding to $r\lesssim 10^{-3}$.

However, it is also recognized that for the next generation of very sensitive CMB polarization experiments the major limitation will come from a non-trivial combination of astrophysical and instrumental systematics. Depending on how they originate, systematics can be associated with the Galactic and extragalactic  foregrounds whose emission is orders of magnitude above the sensitivity levels that are expected to be attained by the experiments, with the non-ideal performance of the instruments or, more importantly, with a combination of both. In relation to the performance of the experiment it is important to establish the requirements 
in relation to key instrumental quantities, such as the optical beam and noise models, pointing accuracy, polarization efficiency and angle, etc. These requirements represent, therefore, the main goals to be accomplished by means of the instrumental calibration methodologies that are being developed during the last years \cite{Cal_Sensors,CubeSat15,Self-Cal,Nati2017,Aumont2019,Keating2013,Minami2019,Xu2020,Minami2020a, Minami2020b,Abitbol2021,CS21}. 

In this paper we focus on the requirements related to the polarization angle accuracy \cite{IF_Pol, delaHoz2021}. We estimate both absolute and relative polarization angle calibration requirements. In fact, knowledge and measurement of the orientation of each detector with respect to an absolute observational reference frame is required. This can be achieved by considering the orientation of the focal planes in the experiment and inter-calibration relative orientation between different groups of detectors, hence local reference frames associated classified by frequency, focal plane wafer, etc.. Such a functional decomposition of the problem has the utility of enabling common alignment errors to be collectively treated~\citep[see, e.g.,][]{Koopman2016}.

In this sense, although the main classification of detectors is given by their frequency, in order to link the reported requirements with component separation methods related to the foreground signals, other classifications have been also taken into account, considering possible correlations between frequency and focal plane or even frequency and detector wafers. 

On the other hand, it is important to note that this work is focused on the estimation of requirements, but a methodology to meet those requirements is not proposed. This later subject is studied in \citep{IF_Pol} and \citep{delaHoz2021}.

In Section~\ref{sec:method} we describe the methodology to determine the requirements for a given experimental configuration. We discuss how detectors' systematics bias the $r$ parameter in Section~\ref{sec:method_c}. In Section~\ref{sec:results} we apply our methodology to the case of LiteBIRD to assess the accuracy on the polarization angle matching the LiteBIRD requirements, in particular on $r$ accuracy. Finally, conclusions are given in Section~\ref{sec:conclusions}.

\section{Methodology}
\label{sec:method}
We present a method to determine the polarization angle calibration requirements given an expected signal level of the primordial gravitational waves.

The CMB polarization signals in terms of the Stokes parameters $(Q, U)$ can be estimated as a linear combination of the contributions at the different frequency elements of the experiment as:
\begin{equation}
\label{eq:CMBpol}
\left({\begin{array}{c} \hat{Q} \\ \hat{U}  \end{array}}\right) (p) = 
\sum_{\nu=1}^n \omega_\nu \left({\begin{array}{c} Q_\nu \\ U_\nu  \end{array}}\right) (p),
\end{equation}
where $\nu$ identifies each of the $n$ different frequency elements, $p$ identifies a pixel of the map, $\omega_\nu$ is the weight for the element at frequency $\nu$ and $(Q_\nu,U_\nu)$ are the Stokes parameters from the polarization data at frequency $\nu$.
Equivalently, the two CMB spherical harmonics coefficients in polarization, $\hat{e}_{\ell m}$ and $\hat{b}_{\ell m}$ modes (\citep{Zaldarriaga1997, Kam96}), are given by:
\begin{equation}
\label{eq:CMBpol_modes}
\left({\begin{array}{c} \hat{e}_{\ell m} \\ \hat{b}_{\ell m}  \end{array}}\right) =
\sum_{\nu=1}^n \omega_\nu \left({\begin{array}{c} e^\nu_{\ell m} \\ b^\nu_{\ell m}  \end{array}}\right).
\end{equation}
\begin{figure}
    \centering
    \includegraphics[width=\textwidth]{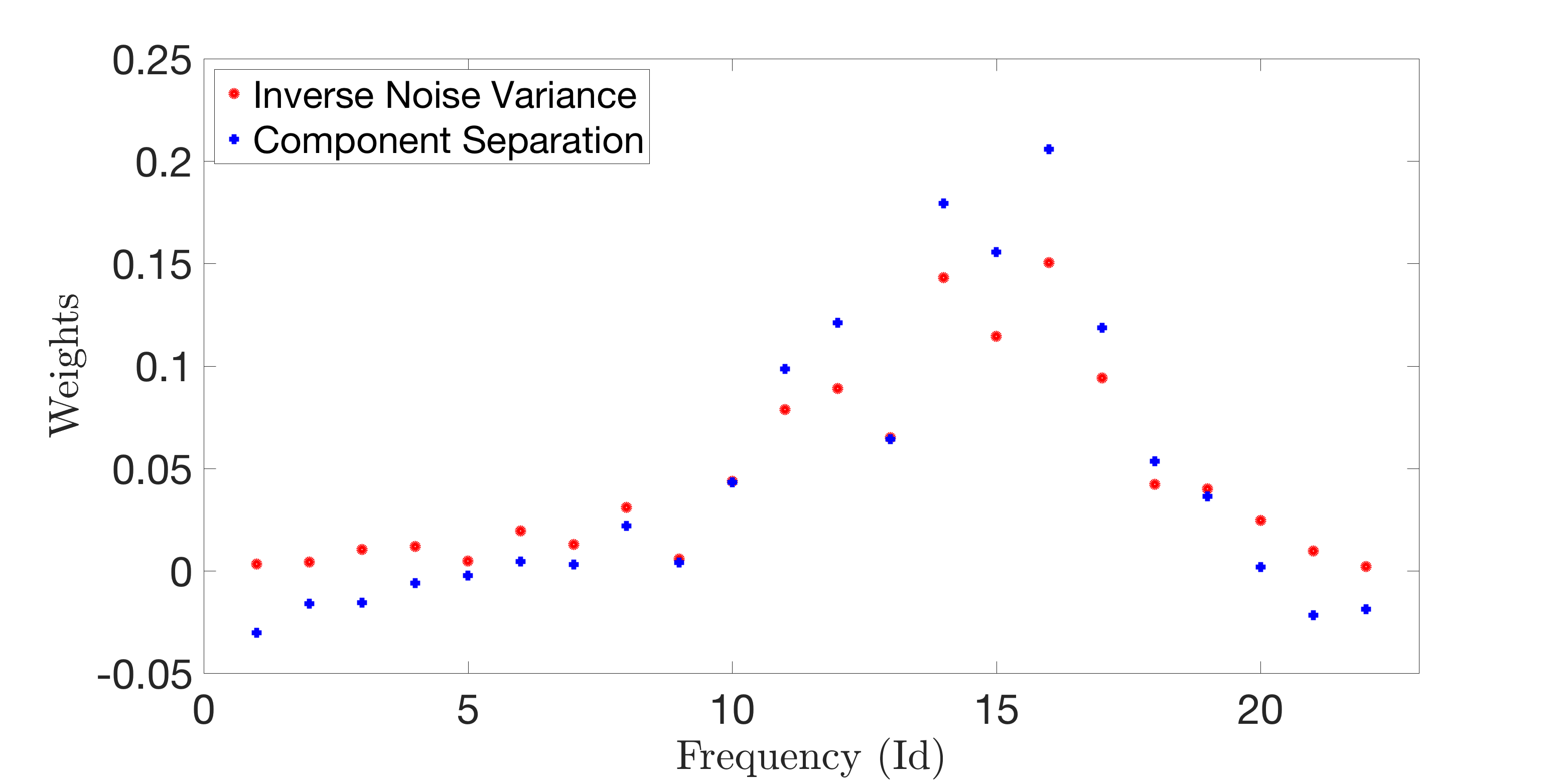}
    \caption{Inverse noise variance weights compared to the weights obtained with the \texttt{FGBuster} parametric methodology developed in \cite{Errard2019}. The figure refers to the weights obtained for the 22 frequency elements of the LiteBIRD instruments, as listed in Table~\ref{instrument}.}
    \label{weights}
\end{figure}

Component separation methods based on a linear combination of frequency data, generically known as ILC (Internal Linear Combination) methods, estimate the weights $w_\nu$ by imposing that the resulting CMB polarization map has minimum variance, following the formalism of \cite{Fernandez-Cobos2016} (\texttt{PILC} and \texttt{PRILC}). Notice that these linear combination methods assume the following constraint for the weights:  $\sum_{i=1}^n w_i=1$. A particularly interesting case of the ILC is the Internal Template Fitting~\citep[e.g., \texttt{SEVEM},][]{Fernandez-Cobos2012}, which minimizes the variance of the residuals after the subtraction of a linear combination of internal foreground templates from the original frequency map.

On the other hand, a parametric method like \texttt{FGBuster} described in \cite{Errard2019} and used as a baseline for LiteBIRD~\cite{PTEP2021}, is expected to provide $w_\nu$ values close to the weights that would be obtained through a inverse noise variance weighting scheme\footnote{J. Errard, private communication.}  (see Fig.~\ref{weights}). Hereinafter, we will assume this type of weights based on the inverse noise variance for deriving the results given in Section~\ref{sec:results}.   
Since we are setting requirements also on the relative polarization angle between different detector sets (e.g., a subset of the detectors belonging to a given wafer), hereinafter (unless explicitly stated) $n$ will indicate the total number of detector sets.

When the polarization axes are rotated by an angle $\alpha_\nu$ the Stokes parameters $Q,U$ are transformed to the rotated ones $Q^{\mathrm{rot}},U^{\mathrm{rot}}$ as follows:
\begin{eqnarray}
\label{eq:rotation}
Q^{\nu,\mathrm{rot}}(p) & = &  \cos2\alpha_\nu \, Q^{\nu}(p)-\sin2\alpha_\nu \, U^{\nu}(p) \\
U^{\nu,\mathrm{rot}}(p)  & = & \sin2\alpha_\nu \, Q^{\nu}(p)+\cos2\alpha_\nu \, U^{\nu}(p)   \nonumber.
\end{eqnarray}
Notice that here we assume that the only source for mixing E- and B-modes are only due to this polarization angle due to instrumental effects, and we do not consider other astrophysical (e.g., foregrounds) or cosmological (e.g., birefringence) effects. Assuming a uniform rotation over the sky, the polarization modes $e_{\ell m}$ and $b_{\ell m}$ are also transformed in the same way:
\begin{eqnarray}
\label{eq:rotation_modes}
e^{\nu,\mathrm{rot}}_{\ell m} & = & \cos2\alpha_\nu \, e_{\ell m}^{\nu}(p)-\sin2\alpha_\nu \, b_{\ell m}^{\nu}(p)  \\
b_{\ell m}^{\nu,\mathrm{rot}} & = & \sin2\alpha_\nu \, e_{\ell m}^{\nu}(p)+\cos2\alpha_\nu \, b_{\ell m}^{\nu}(p)  \nonumber .
\end{eqnarray}
The CMB polarization signal, estimated from the $n$ frequency elements, is modified when the polarization axes of each element $\nu$ are rotated by an angle $\alpha_\nu$. Following eqs.(\ref{eq:CMBpol_modes}, \ref{eq:rotation_modes}), we get:
\begin{eqnarray}
\label{eq:CMB_rot}
\hat{e}_{\ell m} & = & e_{\ell m}\sum_{\nu=1}^n \omega_\nu \cos2\alpha_\nu - b_{\ell m}\sum_{\nu=1}^n \omega_\nu \sin2\alpha_\nu \\
\hat{b}_{\ell m} & = & e_{\ell m}\sum_{\nu=1}^n \omega_\nu \sin2\alpha_\nu + b_{\ell m}\sum_{\nu=1}^n \omega_\nu\cos2\alpha_\nu \nonumber ,
\end{eqnarray}
The corresponding change in the B-mode CMB angular power spectrum, $B_{\ell}$, would be:
\begin{equation}
\label{eq:CMB_power}
    \hat{B}_{\ell}=E_{\ell}\left(\sum_{\nu=1}^n w_\nu \sin2\alpha_\nu \right)^2+B_{\ell}\left(\sum_{\nu=1}^n w_\nu \cos2\alpha_\nu\right)^2 \ .
\end{equation}
Let's now consider the error that such rotations will induce in the tensor-to-scalar parameter $r$. Assuming a Gaussian likelihood approximation (see Appendix~\ref{sec:Appendix_A}), for the B-mode CMB angular power spectrum, that error, $\delta r$, is given by the following expression:
\begin{equation}
    \label{eq:delta_r}
    \delta r = \left[ \sum_{\ell =2}^{\ell _{\mathrm{max}}}{\frac{\Delta B_{\ell} B_{\ell}^\mathrm{fid}}{\mathrm{Var}(B_{\ell})}}\right] \left[ \sum_{\ell =2}^{\ell _{\mathrm{max}}}{\frac{(B_{\ell}^\mathrm{fid})^2}{\mathrm{Var}(B_{\ell})}} \right]^{-1}\\.
\end{equation}
In this equation, $B_{\ell}^\mathrm{fid}$ is the B-mode CMB angular power spectrum due to inflationary gravitational waves and for the fiducial $\Lambda$CDM model with $r=1$. $\Delta B_{\ell}$ is the shifted B-mode CMB angular power spectrum resulting after subtracting, from the estimated B-mode CMB angular power spectrum $\hat{B}_{\ell}$, the known components: the $B_{\ell}^\mathrm{fid}$, lensing ($L_{\ell}$), residual foregrounds ($R_{\ell}^E$ and $R_{\ell}^B$), and the effective noise $N_{\ell}^\mathrm{eff}$ angular power spectra. Therefore, the final estimated B-mode CMB angular power spectrum is given by:
\begin{equation}
    \label{eq:BB_obs}
    \hat{B}_{\ell}=\left(rB_{\ell}^\mathrm{fid}+L_{\ell}+R_{\ell}^B\right)\Sigma_{cos}+(E_{\ell}+R_{\ell}^E)\Sigma_{sin}+N_{\ell}^\mathrm{eff}\ ,
\end{equation}
where $E_{\ell}$ is the fiducial E-mode CMB angular power spectrum, and  $R_{\ell}^E$ and $R_{\ell}^B$ are the angular power spectra of the foreground residuals for the E- and B-mode, respectively. Generically, these residuals are expected to be of the same order of magnitude in amplitude, with $R_{\ell}^E<<E_{\ell}$ and $R_{\ell}^B\sim L_{\ell}$. The terms $\Sigma_{cos}$ and $\Sigma_{sin}$ account for the polarization angle offsets (that are assumed to be random variables of zero mean) of all the frequency detector sets and will be given below. The effective noise power spectrum, $N_{\ell}^\mathrm{eff}$, is obtained as a weighted combination of the noise of all the frequency elements,
\begin{equation}
    \label{eq:noise_eff}
    N_{\ell}^\mathrm{eff}=\sum_{\nu=1}^n w_{\nu}^2N_{\ell}^{\nu} \left(b_{\ell}^{\nu}\right)^{-2}\ .
\end{equation}
where $b_{\ell}^{\nu}$, stands for the beam window function associated with the $\nu$ data. As mentioned above, the shifted B-mode CMB angular power spectrum, $\Delta B_{\ell}$ (or equivalently the bias in the B-mode CMB angular power spectrum due to the presence of polarization angle offsets), is obtained by subtracting the known contributions from the observed spectrum $\hat{B}_{\ell}$:
\begin{equation}
    \label{eq:BB_bias}
    \Delta B_{\ell}=\left(rB_{\ell}^\mathrm{fid}+L_{\ell}+R_{\ell}^B\right) (\Sigma_{cos} -1 ) + (E_{\ell}+R_{\ell}^E)\Sigma_{sin} \ . 
\end{equation}
Obviously, those contributions can be removed at the power spectrum level but they have an effect in the cosmic variance (e.g., here we do not attempt to do any delensing):
\begin{equation}
    \label{eq:cosmic_var}
    \mathrm{Var}(B_{\ell})=\frac{2}{f_{\mathrm{sky}}(2l+1)}\hat{B}_{\ell}^2,
\end{equation}
where $f_{\mathrm{sky}}$, the fraction of sky used, is accounting for the sampling variance. As commented above, the $\Sigma_{\cos}$ and $\Sigma_{\sin}$ account for the impact of the polarization angle offsets $\alpha_\nu$,
\begin{equation}
    \label{eq:Sigmas}
    \Sigma_{\sin}=\left(\sum_{\nu=1}^n w_\nu \sin2\alpha_\nu \right)^2 \ \ , \ \ 
    \Sigma_{\cos}=
    \left(\sum_{\nu=1}^n w_\nu \cos2\alpha_\nu\right)^2 \ .
\end{equation}
From the above equations (\ref{eq:delta_r},\ref{eq:BB_bias},\ref{eq:Sigmas}) it is clear that in the limit where all the offsets become negligible, $\alpha_\nu \to 0$, then $\Sigma_{\sin}=0$ and $\Sigma_{\cos}=1$. This implies that there is no error in the B-mode CMB angular power spectrum, $\Delta B_{\ell}=0$, or in the $r$ parameter, $\delta r=0$, as expected. Also, notice that random (coherent) errors in the detector angles will tend to decrease (increase) $\Delta B_{\ell}$ and $\delta r$, as it is shown in Section~\ref{sec:results}.

Since the next generation of CMB experiments are expected to provide very sensitive measurements of the CMB B-modes, instrumental errors associated with the polarization angles above a few degrees would prevent them from achieving their scientific objectives. Therefore,  it is worth to consider the small angle approximation in Eq.~\ref{eq:Sigmas}:
\begin{equation}
    \label{eq:Sigmas_small}
    \Sigma_{\sin}\approx 4\left(\sum_{\nu=1}^n w_\nu \alpha_\nu \right)^2 \ \ , \ \ 
    \Sigma_{\cos}\approx1-4
    \sum_{\nu=1}^n w_\nu \alpha_\nu^2 \ ,
\end{equation}
and the error in $r$ takes the following simple expression:
\begin{equation}
    \label{eq:delta_r_small}
    \delta r = 4A
    \left(\sum_{\nu=1}^n w_\nu \alpha_\nu \right)^2\ \ ,
\end{equation}
where we have introduced the factor $A$ that is given by:
\begin{equation}
 \label{eq:A_factor}   
 A = 
 \left[\sum_{\ell =2}^{\ell _{\mathrm{max}}}{\frac{(E_{\ell}+R_{\ell}^E)B_{\ell}^\mathrm{fid}}{\mathrm{Var}(B_{\ell})}}\right]    \left[ \sum_{\ell =2}^{\ell _{\mathrm{max}}}    {\frac{(B_{\ell}^\mathrm{fid})^2}{\mathrm{Var}(B_{\ell})}}   \right]^{-1}.
 \end{equation}
\begin{figure}
    \centering
    \includegraphics[width=\textwidth]{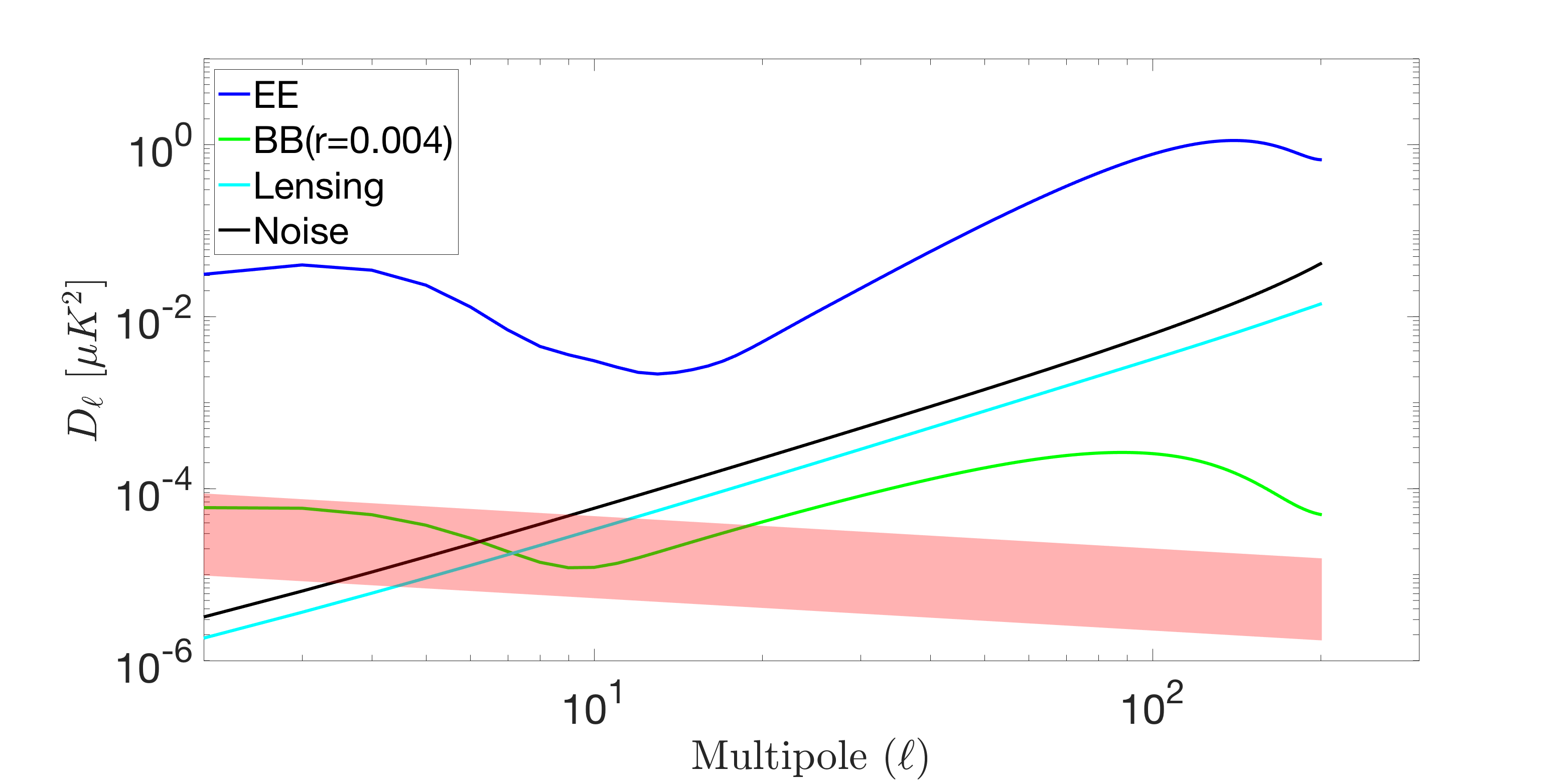}
    \caption{Angular power spectra of the different signals and of the estimate of the foreground residuals. The latter is taken from ~\cite{Diego-Palazuelos2020}, and it is shown as a band that accounts for a typical foreground residual given by $R_{\ell}^B = 2\times 10^{-2}\ell^{-2.29}$, as estimated by \cite{Errard2016} for the LiteBIRD experiment. The band expands over a range of a factor 1/3 and 3 times this function. For the noise, the effective LiteBIRD amplitude is plotted as reference}
    \label{residuals}
\end{figure}

Notice that we have also considered that $rB_{\ell}^\mathrm{fid}+L_{\ell}+R_{\ell}^B<<E_{\ell}$ in order to neglect the second order term in the approximation for $\Sigma_{cos}$ in Eq.~\ref{eq:Sigmas_small}. Whereas the current knowledge of the cosmological parameters of the standard model allows one to have a good enough determination of the cosmological signals (i.e., CMB and lensing), it is not the case for the expected amount of residual foregrounds. For the later, we assume that both contribution to E- and B-modes are of the same order ($R_{\ell}^E\approx R_{\ell}^B$), and we adopt the estimations provided in (\cite{Diego-Palazuelos2020}, \cite{Errard2016}).
Fig.~\ref{residuals} shows the amplitude of the different  signals and the expected foreground residuals at the power spectrum level. The latter is represented by a band that accounts for an uncertainty of a factor $3$ from the estimate just quoted. As it can be seen, the B-mode lensing, residual foregrounds and also the primordial B-mode signals (even if we would allow for values of $r$ equal to current upper limits for the latter) are much smaller than the CMB E-mode, what justifies neglecting the second order term in $\Sigma_{cos}$. In addition, the residual foregrounds can potentially impact the polarization angle requirements through the variance term in Eq.~\ref{eq:delta_r_small}.

However, we have checked that in practice, for the typical foreground residuals expected from optimal component separation methods (see Fig.~\ref{residuals}), this impact implies a less restrictive requirement for the polarization angle. We will here adopt the typical residual expected for an experiment like LiteBIRD, as estimated in~\cite{Errard2016}. We will discuss their impact for different amplitudes of the residuals in Section~\ref{sec:results}, when we compute the requirements for the specific case of the LiteBIRD mission.  Eq.~\ref{eq:delta_r_small} is a general expression that only depends on the polarization angle mismatch at each frequency element, $\alpha_\nu$, and the weight that each element has to build the final CMB map, $w_\nu$.

\section{Correlation among detectors and bias in $r$}
\label{sec:method_c}

Correlation of the polarization angle offsets across different detectors may come from several systems of the experiment, such as the focal plane, optical components or the platform. Such instrumental systematic effects affect different sets of detectors at the same time, but in general at different levels.
 For instance, a mismatch between the absolute polarization angle reference of the experiment (or, in the case of a space mission, the Service Module reference) and the sky reference will affect all the instruments and, therefore, all the detectors. Also, the presence of non-idealities in the reflecting mirrors or refracting lenses, or of some wobbling in the Half Wave Plate, will create common artifacts in all the detectors of the illuminated focal plane that, in particular, may translate in possible detector offsets that will be then correlated. Detectors belonging to the same frequency element may also be affected by similar band-response systematics, or, if they are in the same wafer, they may share miss-calibrations related to that wafer. 

{We present here a formalism that can be applied to different levels of complexity to define the polarization angle offset. For the shake of simplicity, we will make our general considerations for the case in which we have a polarization angle requirement for $n$ frequency elements. Those could be the actual frequency bands of a given experiment, or any other division of detectors by frequency sets. Latterly, in Section~\ref{sec:method_D} we will consider the case of relative polarization angle requirements at instrument or telescope level, plus an absolute one at satellite level. Finally, in Section~\ref{sec:detectors}, we will explain how to propagate the requirements imposed at the frequency elements levels to a more fainter detector level division.

Denoting by $C$ the matrix that defines the correlations among polarization angle offsets, $\alpha_\nu$, then we have
\begin{equation}
    \langle \alpha_{\nu_1} \alpha_{\nu_2} \rangle \equiv C_{\nu_1\nu_2}=\rho_{\nu_1\nu_2}\sigma_{\nu_1}\sigma_{\nu_2} \ \ ,
\end{equation}
where we have introduced the correlation coefficient  $\rho_{\nu_1\nu_2}$ corresponding to frequency elements $\nu_1$ and $\nu_2$, and $\sigma_\nu$ denotes the dispersion of the polarization angle at frequency element $\nu$. As we will show below, $\sigma_\nu$  represents the uncertainty of the polarization angle for a given frequency element. We want to evaluate some requirements in order to limit its impact on the parameter $r$. Let us now consider the expected value in Eq.~\ref{eq:delta_r_small} to get the expected bias in $r$, $\langle \delta r\rangle$, that a set of uncertainties in the polarization angle, together with their correlations represented by $\sigma_\nu$ and $C_{\nu_1\nu_2}$ respectively, will produce:
\begin{equation}
    \label{eq:delta_r_average}
    \langle \delta r \rangle = 4 A
    \sum_{\nu_1,\nu_2=1}^n C_{\nu_1 \nu_2} w_{\nu_1} w_{\nu_2}\ \ .
\end{equation}
From this equation we see that, given a limiting value for the expected bias in $r$, there are many different combinations of the set of $n$  uncertainties ($\sigma_\nu$) that can lead to the same $\langle\delta r\rangle$. A natural approach to solve this problem, and determine the accuracy with which we need to know the polarization angles ($\sigma_\nu$), is to assume that all the terms in the sum of Eq.~\ref{eq:delta_r_small} add evenly, i.e., $\sigma_\nu=c|w_{\nu}|^{-1}$, where $c$ is a constant\footnote{Although the modulus of the weights need to be considered to ensure that $\sigma_\nu$ is positive, in practice for the inverse-variance weights assumed by default, which are obviously positive, we will not consider the modulus for the rest of the paper. The generalization of the formalism to the case of negative weights is given in Appendix~\ref{sec:Appendix_B}}. With this assumption, an expression can be obtained to provide the requirements on $\sigma_\nu$:
\begin{equation}
    \label{eq:delta_r_assumption}
    \langle \delta r \rangle = 4c^2 A
    \sum_{\nu_1,\nu_2=1}^n \rho_{\nu_1 \nu_2}\ \ .
\end{equation}
Therefore, the entire challenge is reduced to determination of the constant, $c$\footnote{Notice that if anti-correlations are considered, the result of the sum should be taken in absolute value}.

Let us now consider two extreme cases for the correlation matrix. In the first case, all the $n$ $\alpha_\nu$ are fully correlated, therefore, the matrix $\rho$ will have all its elements equal to unity, i.e. $\rho_{\nu_1\nu_2}=1$ for any pair of frequency elements ($\nu_1$,$\nu_2$). Then the expected bias for $r$ is in this case:
\begin{equation}
    \label{eq:delta_r_corr}
    \langle \delta r \rangle = 4n^2c^2 A 
    \ \ .
\end{equation}
As expected, this is the case where the most stringent requirements are obtained for the uncertainties in the polarization angles, and any other alternative scheme (i.e., with some $\rho_{\nu_1\nu_2}<1$) will provide weaker requirements. Notice that the same behaviour would be obtained for fully anti-correlations. In the second case we consider an "ideal" experiment where all the polarization angles are uncorrelated, i.e., $\rho_{\nu_1\nu_2}=0$ for $\nu_1\neq \nu_2$. We refer to this case as "ideal", because in this scenario there should not be any kind of systematics that can correlate the angles. The expected bias in $r$ would be now:
\begin{equation}
    \label{eq:delta_r_random}
    \langle \delta r \rangle = 4nc^2 A 
    \ \ .
\end{equation}
As expected, in this case the requirements (given by the constant $c$) are $\sqrt{n}$ weaker than for the case of fully correlated elements.

Finally, let us remark that one could also consider the possibility of constraining the polarization angle through limiting the standard deviation, instead of the offset, of the $r$ parameter. We have explored this option, and found that it provides less stringent constraints on the polarization angles.

\subsection{Relative and absolute angles}
\label{sec:method_D}

Let us now consider an experiment with $m$ telescopes (e.g. $m=3$ for the case of LiteBIRD) and their corresponding focal planes that will be denoted by $F_1, ..., F_m$.  In this case, in addition to the detector or element reference frame we can identify other reference frames for the polarization angles that can be classified as absolute and relative. 

For the absolute angles we consider a global angle associated with a global reference frame common to all the focal planes, and additional ones associated with the corresponding $m$ focal plane reference frames. For the relative angles we can consider the ones associated with $n$ different frequency elements (e.g., frequency bands, wafers, or even to each individual detector). The goal is to impose accuracy requirements on possible offsets of the global polarization angle and of the focal planes relative to the global one (absolute angles); and also the possible polarization angle offsets of each element included in a given focal plane with respect to its corresponding focal plane reference frame (relative angles).    

The requirements corresponding to the relative angles can be estimated with the expression given by Eq.~\ref{eq:delta_r_assumption}. For the absolute angles the requirements can be estimated following similar steps to the ones used for the relative angles as follows. From the error induced on $r$, $\delta r$, in the small angle approximation, Eq.~\ref{eq:delta_r_small}, the contribution to $\delta r$ from the global offset $\alpha_g$ and the additional $m$ offsets corresponding to each of the focal planes $\alpha_{F1}, ..., \alpha_{Fm}$, is given by:
\begin{eqnarray}
    \label{eq:delta_r_small_abs}
    \delta r  & = & 4 A 
    \Bigg[
    \sum_{i=1}^{n_{F1}} {w_i   (\alpha_g+\alpha_{F1})} + \sum_{i=n_{F1}+1}^{n_{F1}+n_{F2}}{ w_i (\alpha_g+\alpha_{F2})}+ ... + \nonumber \\
    & & \sum_{i=n_{F1} + ... + n_{F(m-1)}+1}^{n}
    {w_i (\alpha_g+\alpha_{Fm})}
    \Bigg]^2 \ \,
\end{eqnarray}
where $n_{Fi}$ is the number of frequency elements of the focal plane $Fi$, and $n=n_{F1}+ ... +n_{Fm}$ is the total number of frequency elements in all the focal planes. Taking the average in the previous equation, we obtain:
\begin{eqnarray}
    \label{eq:delta_r__abs_average}
    \langle \delta r \rangle & = & 4 A 
    \Bigg[
    \sigma_g^2+\sum_{i=1}^m{w_{Fi}^2 \sigma_{Fi}^2} + 2\sum_{i=1}^m{w_{Fi} C_{g{Fi}}} + 2\sum_{i\neq j=1}^{m} {w_{Fi} w_{Fj} C_{{Fi}{Fj}}} \Bigg] \ \ , 
\end{eqnarray}
where the weight associated with the focal plane $Fi$ is $w_{Fi}=\sum_{i=1}^{n_{Fi}} w_i$, and $C_{g{Fj}}$ and $C_{{Fi}{Fj}}$ denote the polarization angle correlations between the global and the individual focal plane, and among individual focal plane, respectively. We have used also the fact that $w_g=\sum_{i=1}^n w_i=1$.

As we did for the derivation of the bias $\delta r$ in the case of the relative angles, we also assume that all the terms in the sum of $\delta r$ add evenly on average, i.e., $\sigma_g=c$ and $\sigma_{Fi}=c|w_{Fi}|^{-1}$ with $c$ constant.
We adopted this choice so that all the actors involved in the bias contribute in a similar way.
Considering this, and replacing the correlations by the corresponding correlation coefficients, we finally have:
\begin{equation}
    \label{eq:delta_r__abs_corr}
    \langle \delta r \rangle  =  4c^2 A 
   \Bigg[(m+1)+2\sum_{i=1}^m \rho_{g{Fi}} + 2\sum_{i\neq j=1}^{m} \rho_{{Fi}{Fj}} \Bigg] .
\end{equation}

\subsection{Relative angles at detector level}
\label{sec:detectors}
\label{sec:correlation_det}

Finally, it is worth also discussing possible requirements on the polarization angle established at the single detector level. Notice that for experiments like LiteBIRD, that would imply to extend the methodology to thousands of detectors. However, it is
possible to find a shortcut for the requirements at the detector level, starting from those already found at a higher level (i.e., at the frequency element level).

To show that, let us rewrite Eq.~\ref{eq:delta_r_assumption}, now assuming that it relates the bias on $r$ as a function of the correlation coefficient matrix among polarization angle offsets for each frequency element (for consistency in the notation, in this subsection we will denote $n_\mathrm{f} \equiv n$ as the number of frequency elements):
\begin{equation}
    \label{eq:delta_r_freq_level}
    \langle \delta r \rangle_{\mathrm{freq}} = 4c_\mathrm{f}^2 A
    \sum_{\nu_1,\nu_2=1}^{n_\mathrm{f}} \rho_{\nu_1 \nu_2} \equiv 4c_\mathrm{f}^2 A \mathbf{\rho_\mathrm{f}} .
\end{equation}
where $c_\mathrm{f}$ is the same constant $c$ as in Eq.~\ref{eq:delta_r_assumption}, but including now the sub-index $\mathrm{f}$ to emphasize that this is obtained for a division in frequency elements. Similarly, $\mathbf{\rho_\mathrm{f}}$ represents the correlation structure among frequency elements. 

It is obvious that this bias on $r$ could be also expressed trivially as a function of the correlation coefficient matrix among polarization angle offsets for each detector:
\begin{equation}
    \label{eq:delta_r_det_level}
    \langle \delta r \rangle_{\mathrm{det}} = 4c_\mathrm{d}^2 A
    \sum_{\nu_1,\nu_2=1}^{n_\mathrm{d}} \rho_{\nu_1 \nu_2} \equiv 4c_\mathrm{d}^2 A \mathbf{\rho_\mathrm{d}}\ \,
\end{equation}
where, now, $n_\mathrm{d}$ denotes the number of detectors, $c_\mathrm{d}$ is the constant that should be determined for a given $\langle \delta r \rangle_{\mathrm{det}}$, and $\mathbf{\rho_\mathrm{d}}$ represents the correlation structure among the detectors.

Following the discussion in Section~\ref{sec:method_c}, the constants $c_\mathrm{f}$ and $c_\mathrm{d}$ in Eqs.~\ref{eq:delta_r_freq_level} and~\ref{eq:delta_r_det_level}, relate the uncertainty of the polarization angle with instrumental noise variance, in particular, for frequency elements and detectors, we have:
\begin{eqnarray}
\sigma_{\mathrm{f}} & = & c_\mathrm{f} \sigma_{n_\mathrm{f}}^2 \nonumber\\
\sigma_{\mathrm{d}} & = & c_\mathrm{d} \sigma_{n_\mathrm{d}}^2,
\end{eqnarray}
where $\sigma_{\mathrm{f}}$ and $\sigma_{\mathrm{d}}$
are the uncertainties on the polarization angle for a given frequency element ($\mathrm{f}$), and for a given detector ($\mathrm{d}$), respectively. Finally, $\sigma_{n_\mathrm{f}}^2$ is the instrumental noise variance per frequency element and $\sigma_{n_\mathrm{d}}^2$ is the noise variance per detector. 

Obviously, the bias on $r$ must be the same, independently on whether the requirements are established at frequency or at detector levels. Therefore, one can obtain trivially a general expression for the requirements on the polarization angle for a given detector, as a function of the requirement already established for the frequency elements. In particular, if we pay attention to the uncertainty on the polarization angle for a detector of a given frequency element $\sigma_{\mathrm{d}_\mathrm{f}}$, and taking into account that the number of detectors at that frequency element is given by $n_{\mathrm{d}_\mathrm{f}} = \sigma_{n_\mathrm{d}}^2 / \sigma_{n_\mathrm{f}}^2$, we find, after equaling Eqs.~\ref{eq:delta_r_freq_level} and~\ref{eq:delta_r_det_level}: 
\begin{equation}
    \label{eq:errors_equals}
\sigma_{\mathrm{d}_\mathrm{f}} = \sigma_{\mathrm{f}} n_{\mathrm{d}_\mathrm{f}} \left(\frac{\mathcal{\rho_\mathrm{f}}}{\mathcal{\rho_\mathrm{d}}}\right)^{1/2}.
\end{equation}

Notice that this is a general expression, that can be evaluated for any correlation structure among frequency elements ($\rho_\mathrm{f}$) and detectors ($\rho_\mathrm{d}$). 
Even so, it is interesting to explore in detail some particular examples:
\begin{enumerate}
    \item The case in which all the detectors and, therefore, all the frequency elements, are fully correlated. In this case, $\rho_\mathrm{f} = n_\mathrm{f}^2$ and $\rho_\mathrm{d} = n_\mathrm{d}^2$ and, therefore:
    \begin{equation}
    \label{eq:errors_equals_case1}
\sigma_{\mathrm{d}_\mathrm{f}} = \sigma_{\mathrm{f}} n_{\mathrm{d}_\mathrm{f}} \left(\frac{n_\mathrm{f}}{n_\mathrm{d}}\right),
\end{equation}
\item The case in which all the detectors and, therefore, all the frequency elements, are uncorrelated. In this case, $\rho_\mathrm{f} = n_\mathrm{f}$ and $\rho_\mathrm{d} = n_\mathrm{d}$ and, therefore:
\begin{equation}
    \label{eq:errors_equals_case2}
\sigma_{\mathrm{d}_\mathrm{f}} = \sigma_{\mathrm{f}} n_{\mathrm{d}_\mathrm{f}} \left(\frac{n_\mathrm{f}}{n_\mathrm{d}}\right)^{1/2},
\end{equation}
\item The case in which the frequency elements are uncorrelated, but all the detectors of a given frequency element are fully correlated. In this case, $\rho_\mathrm{f} = n_\mathrm{f}$ and $\rho_\mathrm{d} = \sum_{i}^{n_\mathrm{f}} \left( n_{\mathrm{d}_i} \right)^2$ and, therefore:
    \begin{equation}
    \label{eq:errors_equals_case3}
\sigma_{\mathrm{d}_\mathrm{f}} = \sigma_{\mathrm{f}} n_{\mathrm{d}_\mathrm{f}} \left(\frac{n_\mathrm{f}}{\sum_{i}^{n_\mathrm{f}} \left( n_{\mathrm{d}_i} \right)^2}\right)^{1/2}.
\end{equation}
\end{enumerate}

\section{Requirements for LiteBIRD}
\label{sec:results}
\begin{figure}[ht]
    \centering
    \includegraphics[width=\textwidth]{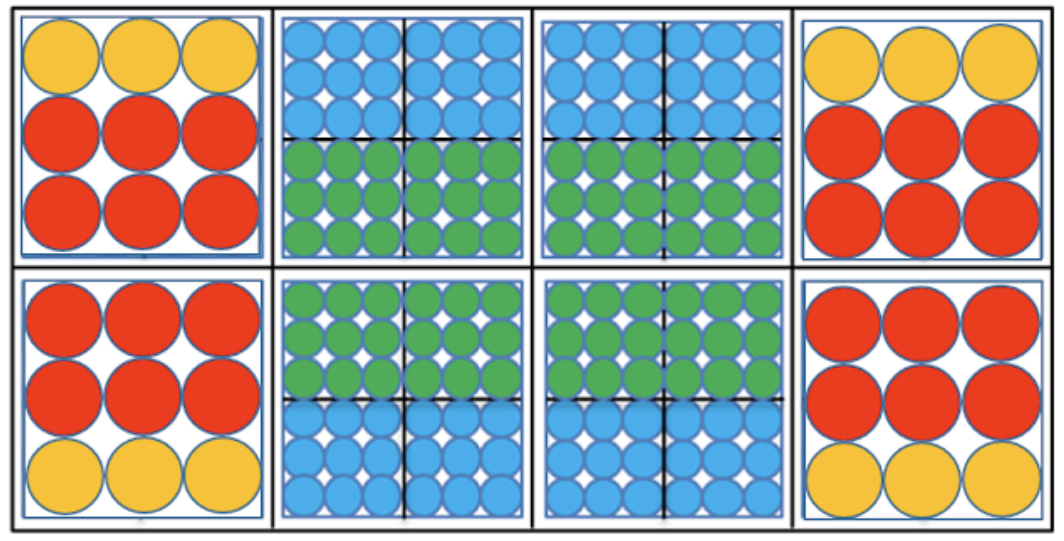}
    \includegraphics[width=0.45\textwidth]{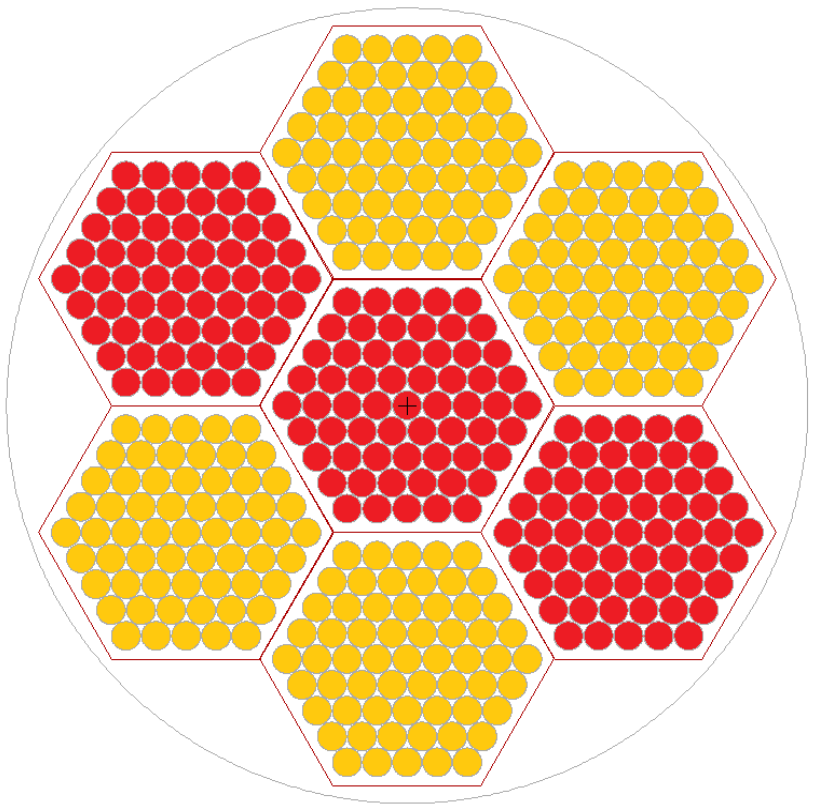}
    \includegraphics[width=0.45\textwidth]{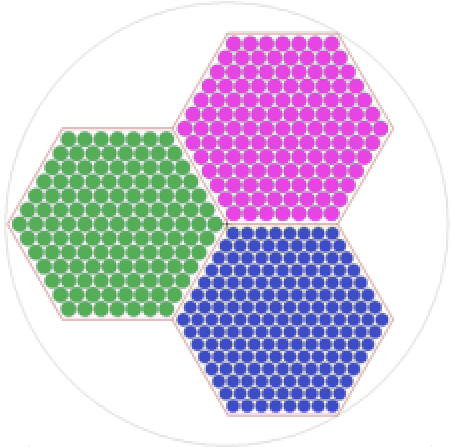}
    \caption{Distribution of detector pixels in the wafers that are inside each of the three telescopes LFT, MFT and HFT. The LFT contains trichroic (operating at three frequency bands) detector pixels distributed among 8 wafers. The MFT contains trichroic and dichroic (two frequency bands) detector pixels distributed among 7 wafers. The HFT contains dichroic and monochroic (operation at one frequency band) detector pixels distributed among 3 wafers.}
    \label{LFT_MFT_HFT}
\end{figure}

The methodology developed in the previous sections to determine the requirements in the accuracy needed in the polarization angle is very general and can be applied in principle to any ground, balloon or space-based experiment. These requirements depend on the characteristics of the experiment, on a chosen upper limit in the bias that this systematics may produce on the tensor-to-scalar ratio $r$, and on the possible correlations that the different polarization offsets can have among themselves. 

As a working example, we consider now the LiteBIRD experiment, a space mission selected by JAXA to be launched in late 2020s \cite{Hazumi2020}, aiming for a sensitivity on $r$ of $\sigma_r = 10^{-3}$ (for $r = 0$), assuming that the error budget is evenly distributed among the systematics, instrumental noise (after foreground removal) and margins. In its present configuration, it contains a total of $22$ frequency elements covering $15$ frequency bands that are distributed in wafers belonging to each one of the three telescopes: the Low Frequency Telescope (LFT), and the Medium (MFT) and High (HFT) Frequency Telescopes (see Fig.~\ref{LFT_MFT_HFT}). The central frequency, FWHM and polarization sensitivity of each frequency element are given in Table~\ref{instrument}. 

\begin{table}
    \centering

         \begin{tabular}{lccccc} 
        Element & ID & Frequency & FWHM  & Pol. sensitivity & Number of\\  
        name & & [GHz] & [arcmin] & $[\mu K$-arcmin] & Bolometers\\
            \noalign{\smallskip}
             \hline
            \noalign{\smallskip}
              LFT\_040GHz & 1 & 40 & 70.5 & 37.42 & 48 \\
              LFT\_050GHz & 2 &  50 &  58.5 & 33.46 & 24 \\
              LFT\_60GHz &  3 & 60 & 51.1 & 21.31 & 48 \\
              LFT\_68GHz\_a  & 4 & 68 & 41.6 & 19.91 & 144 \\
              LFT\_68GHz\_b  & 5 & 68 & 47.1 & 31.77 & 24 \\
              LFT\_78GHz\_a  & 6 &  78 & 36.9 & 15.55 & 144 \\
              LFT\_78GHz\_b  & 7 &  78 & 43.8 & 19.13 & 48 \\ 
              LFT\_89GHz\_a  & 8 &  89 & 33.0 & 12.28 & 144\\
              LFT\_89GHz\_b  & 9 & 89 & 41.5 & 28.77 & 24 \\
              LFT\_100GHz &  10 & 100 & 30.2 & 10.34 & 144 \\
              LFT\_119GHz &  11 & 119 & 26.3 & 7.69 & 144 \\
              LFT\_140GHz &  12 & 140 & 23.7 & 7.25 & 144 \\
              \noalign{\smallskip}
              MFT\_100GHz &  13 & 100 & 37.8 & 8.48 & 366 \\
              MFT\_119GHz &  14 & 119 & 33.6 & 5.70 & 488 \\
              MFT\_140GHz &  15 & 140 & 30.8 & 6.38 & 366\\
              MFT\_166GHz &  16 & 166 & 28.9 & 5.57 & 488 \\
              MFT\_195GHz &  17 & 195 & 28.0 & 7.05 & 366 \\
               \noalign{\smallskip}
            HFT\_195GHz &  18 & 195 & 28.6 & 10.50 & 254\\
            HFT\_235GHz &  19 & 235 & 24.7 & 10.79 & 254\\
            HFT\_280GHz &  20 & 280 & 22.5 & 13.80 & 254\\
            HFT\_337GHz &  21 & 337 & 20.9 & 21.95 & 254\\
            HFT\_402GHz &  22 & 402 & 17.9 & 47.45 & 338 \\
         \end{tabular}
          \caption{Beam resolution, sensitivity and number of detectors corresponding to each of the 22 LiteBIRD frequency elements \cite{PTEP2021}.}
         \label{instrument}
      \end{table}

Following the LiteBIRD strategy, we will assume that the systematic error induced by the use of biased polarization angles in the B-modes analysis can produce, at maximum, a $\delta_r$ of $1\%$ of the total budget assigned to systematics. This total budget of systematics is chosen to be $1/3$ of the overall sensitivity (a quadratic sum accounting for instrumental sensitivity, foreground residuals and overall systematics) on $r$ (i.e., $\delta_r = 5.77\times10^{-6} \equiv \left[10^{-3}/\sqrt{3}\right]\times0.01$). Finally, we assume that only $50\%$ of the sky is used, and that the cosmological model is the one given in \cite{PTEP2021}.

In the following subsections we will provide polarization angle requirements for absolute (Subsection~\ref{sec:results_abs})
and relative angles. For the latter, we will consider 
different schemes for grouping the frequency detectors: at frequency elements (Subsection~\ref{sec:results_rel_freq}) and at wafers on a given telescope (Subsection~\ref{sec:results_rel_wafer}). Starting from any of these two cases (as described in Subsection~\ref{sec:correlation_det}) one could also estimate requirements at detector level. This is discussed in~\ref{sec:results_rel_det}.

\subsection{Absolute angle requirements}
\label{sec:absolute_requirements}
\label{sec:results_abs}

For the absolute angles, the requirements are estimated using Eq.~\ref{eq:delta_r__abs_corr}. The following correlation cases are considered:
\begin{itemize}
    \item Case 1.0: No correlations.
    \item Case 1.1: The four offsets (global and the three ones associated with the focal planes) are fully correlated.
    \item Case 1.2: The global offset is uncorrelated with any of the three focal plane ones, and the latter ones are fully correlated.
    \item Case 1.3: The global offset is fully correlated with any of the three focal plane ones, and the latter are uncorrelated among themselves.
\end{itemize}

Notice that for case 1.2, the correlation coefficients take the values  $\rho_{gF_i}=0$ for any of the three focal planes $F_i$ with $i=1,2,3$, and $\rho_{F_iF_j}=1$ for $i,j=1,2,3$. Also, for case 1.3, we have the reverse situation, with $\rho_{gF_i}=1$ for $i=1,2,3$, and $\rho_{F_iF_j}=0$ for $i,j=1,2,3$.
Therefore, as it can be easily seen from  Eq.~\ref{eq:delta_r__abs_corr}, the two cases provide exactly the same constraints.

In Table~\ref{tab:absreq}, the requirements for the global, LFT, MFT and HFT polarization angles are provided. The requirements are inversely proportional to the inverse of the noise variances  associated, with the global experiment, and with each of the three focal telescopes. Notice that the ratio between the requirements for the Case 1.0 (no correlations) and Case 2.0 (fully correlation) is $\approx 2$. This is an expected result, since for the later we have one effective angle, whereas, for the former, we have 4. Therefore, the sensitivity goes approximately as the square root of the number of elements.

\begin{table}
    \centering
    \begin{tabular}{l|c|c|c|c}
         & \multicolumn{3}{c}{Offset (arcmin)} \\
         Label & Case 1.0 & Case 1.1 & Cases 1.2,\,1.3\\
        \hline
GLB & 3.7 & 1.8 & 2.3  \\
LFT & 11.6 & 5.8 & 7.4  \\
MFT & 6.4 & 3.2 & 4.1  \\
HFT & 30.8 & 15.4 & 19.5
    \end{tabular}
    \caption{Polarization angle requirements for the four absolute offsets for cases 1.0, 1.1 and the two cases 1.2 and 1.3, that provide the same results.}
    \label{tab:absreq}
\end{table}

\subsection{Relative angle requirements at the frequency level}
\label{sec:frequency_requirements}
\label{sec:results_rel_freq}

We will study the requirements on relative angles, following two different schemes for grouping the frequency detectors. We first group the detectors by frequency elements, resulting in a total number of $n=22$ elements and $m=3$ focal planes. Also, the number of elements at each focal plane is $n_{LFT}=12$, $n_{MFT}=5$ and $n_{HFT}=5$, respectively. 

As possible correlation structures, some simplified cases will be considered, including, in particular, the extreme cases of null and full correlations:
\begin{itemize}
\item Case 2.0: The offsets of all the $n$ elements are uncorrelated, except for those in the same telescope focal plane which are fully correlated.
\item Case 2.1: The offsets of all the $n$ elements are fully correlated (strongest constraints).
\item Case 2.2: The offsets of all the $n$ elements are partially correlated, in particular, we chose $\rho_{{\nu_1}{\nu_2}}=0.5$  (for any $\nu_1\neq \nu_2$), except those within the same telescope which are fully correlated.
\item Case 2.3: The offsets of all the $n$ elements are uncorrelated (weakest constraints).
\end{itemize}
The $\rho$ matrix illustrating the correlation among the $22$ elements, for each one of these cases, is given in Fig.~\ref{matrix_correl2}.

\begin{figure}
    \centering
    \includegraphics[width=0.45\textwidth]{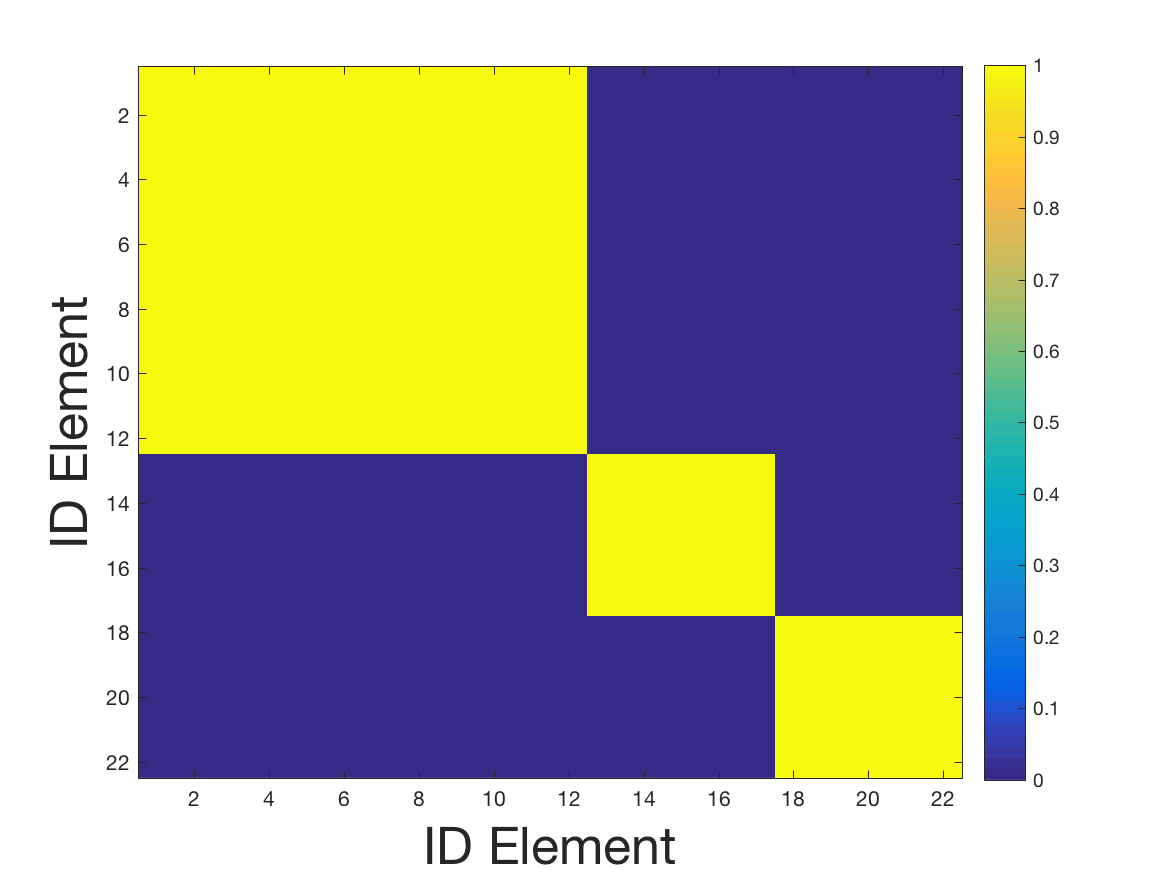}
    \includegraphics[width=0.45\textwidth]{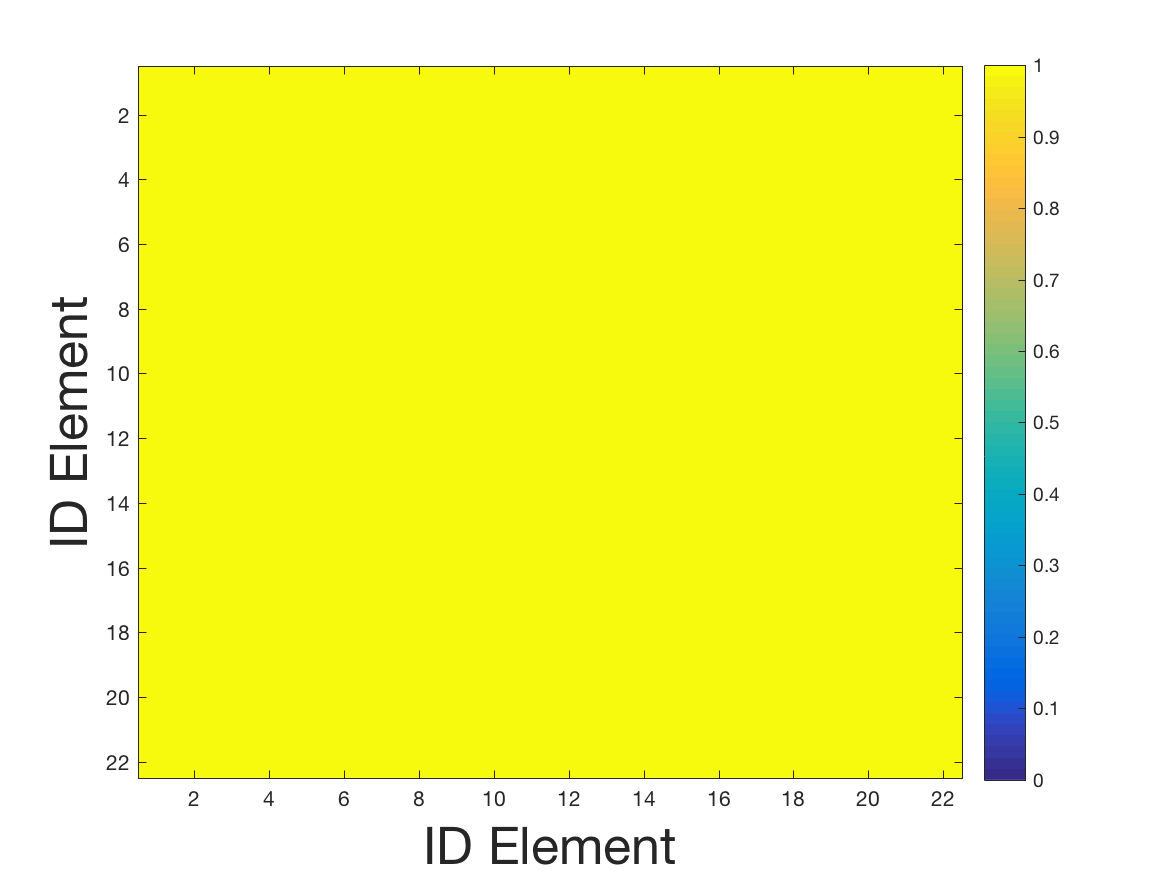}
    \includegraphics[width=0.45\textwidth]{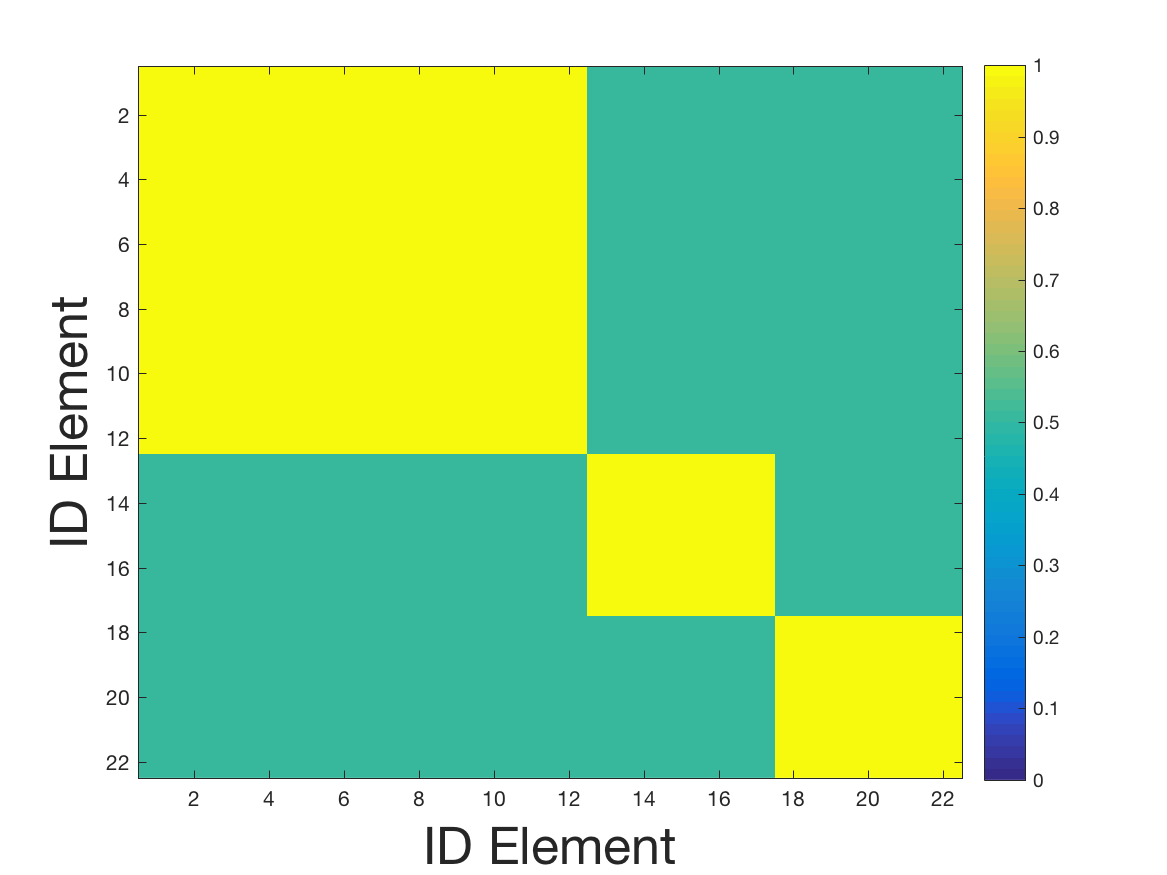}
    \includegraphics[width=0.45\textwidth]{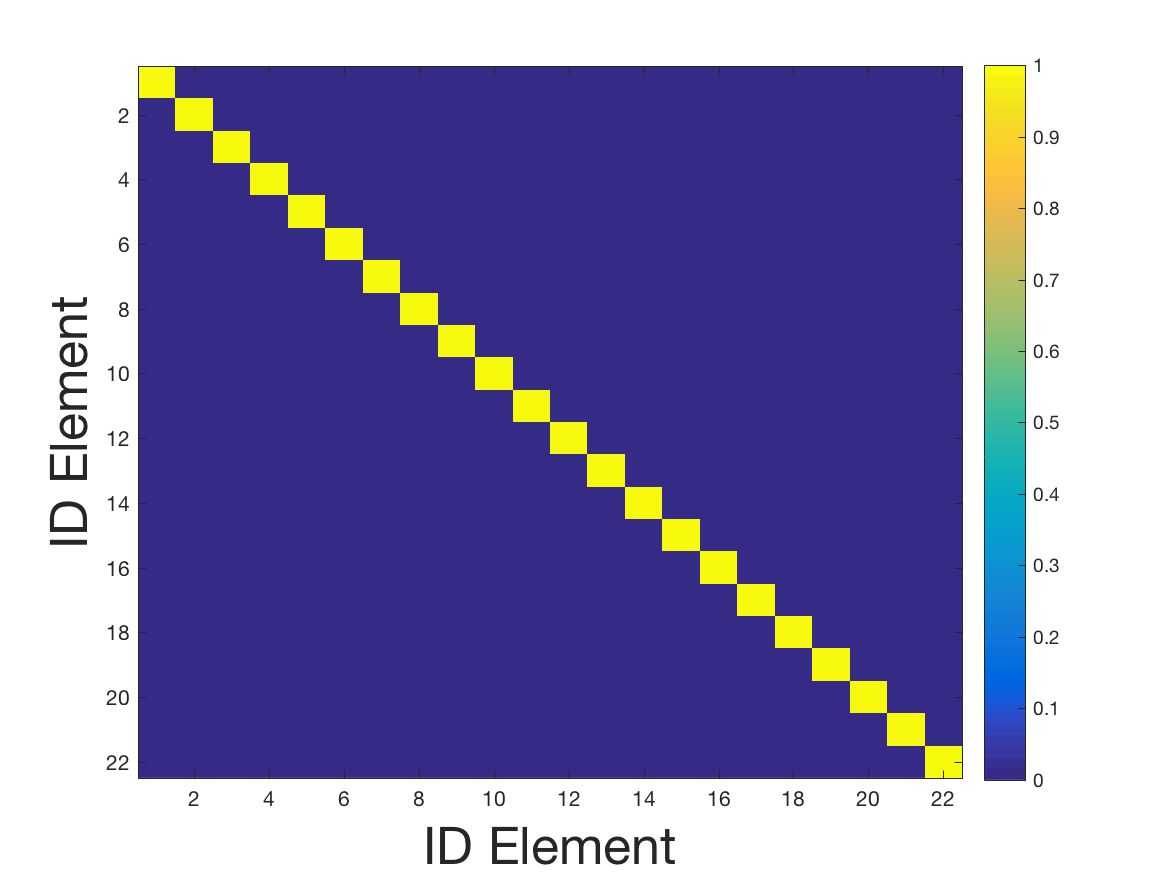}
    \caption{Correlation matrix $\rho$ among the 22 elements within the LFI, MFT and HFT classified by frequency elements. The four different cases 2.0, 2.1, 2.2 and 2.3, considered as illustrative examples, are shown from top-left to bottom-right (see text for details). }
    \label{matrix_correl2}
\end{figure}

In Table~\ref{tab:alpha_relative_frequency} we show the requirements for the detectors of a given frequency element, and for each of the correlation cases considered. As it can be seen, the requirements vary with the sensitivity of the corresponding set of detectors for a given frequency element, and with the correlations among frequency elements. Since the component separation weights are assumed by default to be proportional to the inverse of the noise variance, the requirements are also inversely proportional to that (as was the case of the absolute angles of the previous Subsection). Depending on the frequency element and the correlation case, the requirement for the combined set of detectors can reach very stringent values, even below the arcminute. As previously mentioned, the ratio between the constraints for the Case 2.1 (fully correlated) and Case 2.3 (fully uncorrelated) is approximately the square root of the number of elements, i.e., $\approx 5$.

\begin{table}[]
    \centering
    \begin{tabular}{l|c|c|c|c|c}
     Element  & ID & \multicolumn{4}{c}{$\sigma_\alpha$ (arcmin)} \\
     name & & Case 2.0 & Case 2.1 & Case 2.2 & Case 2.3\\
        \noalign{\smallskip}
            \hline 
              \noalign{\smallskip}
LFT\_040GHz &  1 & 49.8 & 31.5 & 37.7 & 147.8 \\
LFT\_050GHz &  2 & 39.8 & 25.2 & 30.1 & 118.2 \\	
LFT\_060GHz &  3 & 16.1 & 10.2 & 12.2 & 47.9 \\
LFT\_068GHz\_a &  4 & 1.09 & 8.9 & 10.7 & 41.8 \\
LFT\_068GHz\_b &  5 & 35.9 & 22.7 & 27.1 & 106.5 \\
LFT\_078GHz\_a &  6 & 8.6 & 5.4 & 6.5 & 25.5 \\
LFT\_078GHz\_b &  7 & 13.0 & 8.2 & 9.8 & 38.6 \\
LFT\_089GHz\_a &  8 & 5.4 & 3.4 & 4.1 & 15.9 \\
LFT\_089GHz\_b &  9 & 29.4 & 18.6 & 22.3 & 87.4 \\
LFT\_100GHz &  10 & 3.8 & 2.4 & 2.9 & 11.3 \\
LFT\_119GHz &  11 & 2.1 & 1.3 & 1.6 & 6.2 \\
LFT\_140GHz &  12 & 1.8 & 1.2 & 1.4 & 5.6 \\
              \noalign{\smallskip}
MFT\_100GHz &  13 & 2.6 & 1.6 & 1.9 & 7.6 \\
MFT\_119GHz &  14 & 1.2 & 0.7 & 0.9 & 3.4 \\
MFT\_140GHz &  15 & 1.5 & 0.9 & 1.1 & 4.3 \\
MFT\_166GHz &  16 & 1.1 & 0.7 & 0.8 & 3.3 \\
MFT\_195GHz &  17 & 1.8 & 1.1 & 1.3 & 5.3 \\
              \noalign{\smallskip}
HFT\_195GHz &  18 & 3.9 & 2.5 & 3.0 & 11.6 \\
HFT\_235GHz &  19 & 4.1 & 2.6 & 3.1 & 12.3 \\
HFT\_280GHz &  20 & 6.8 & 4.3 & 5.1 & 20.1 \\
HFT\_337GHz &  21 & 17.1 & 10.8 & 13.0 & 50.9 \\
HFT\_402GHz &  22 & 80.0 & 50.7 & 60.5 & 237.6
    \end{tabular}
    \caption{Polarization angle requirements for the 22 frequency elements within the LFT, MFT and HFT, for correlation cases 2.0, 2.1, 2.2, and 2.3 (see text for details).}
    \label{tab:alpha_relative_frequency}
\end{table}

\subsection{Relative angle requirements at the wafer-frequency level}
\label{sec:results_rel_wafer}

As a second illustrative way of classifying the detectors, we consider sets of detectors that have the same frequency and are placed in the same wafer.

As it can be seen in Fig.~\ref{LFT_MFT_HFT}, the division reads:
\begin{itemize}
    \item LFT consists of 8 wafers of 2 different types (4 wafers each). The first type accounts for detectors at 40, 60 and 78GHz (red); and 50, 68 and 89 GHz (yellow). The second type accounts for detectors at 68, 89 and 119 GHz (green); and 78, 100 and 140 GHz (blue).
    \item MFT consists of 7 wafers of 2 different types. The first type accounts for detectors at 100, 140 and 195 GHz (red, 3 wafers). The second type accounts for detectors at 119 and 166 GHz (yellow, 4 wafers).
    \item HFT consists of 3 wafers of 1 type each. The first type accounts for detectors at 195 and 280 GHz (purple). The second type accounts for detectors at 235 and 337 GHz (green). Finally, the third type accounts for detectors at 402 GHz (blue).
\end{itemize}

Therefore, this classification scheme accounts for 70 different frequency detector sets, spread on 7 different types of wafers.
As possible correlations, we will consider simplified cases that, again, in particular, include the extreme cases of null and full correlations:
\begin{itemize}
\item Case 3.0: The offsets of all the $n$ elements are uncorrelated, except for those in the same  telescope focal plane, which are fully correlated.
\item Case 3.1: The offsets of all the $n$ elements are fully correlated (strongest constraints).
\item Case 3.2: The offsets of all the $n$ elements are partially correlated, in particular, we chose $\rho_{{\nu_1}{\nu_2}}=0.5$  (for any $\nu_1$ and $\nu_2$ in the same telescope), except those within the same element which are fully correlated.
\item Case 3.3: The offsets of all the $n$ elements are uncorrelated (weakest constraints).
\end{itemize}
The $\rho$ matrix illustrating the correlation among the $70$ elements, for each one of these cases, is given in Fig.~\ref{matrix_correl3}.

\begin{figure}[ht]
    \centering
    \includegraphics[width=0.45\textwidth]{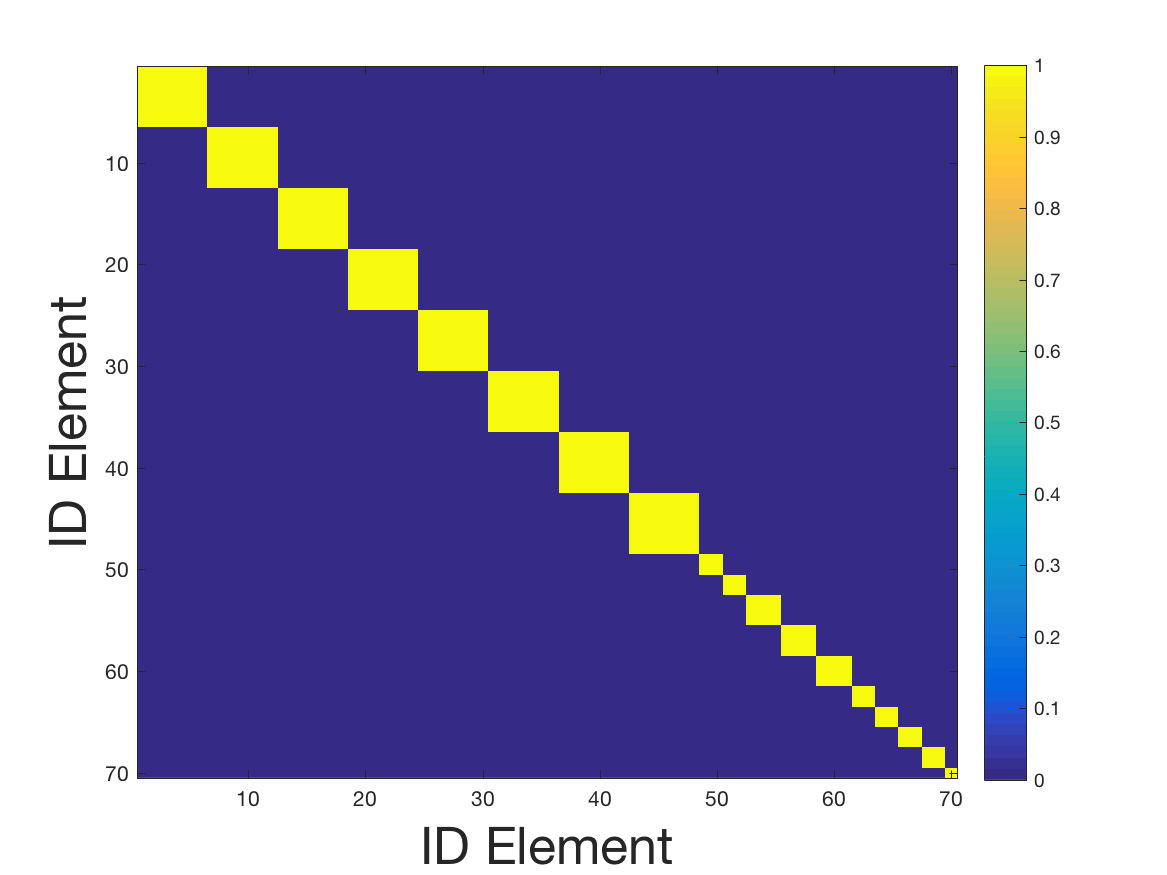}
    \includegraphics[width=0.45\textwidth]{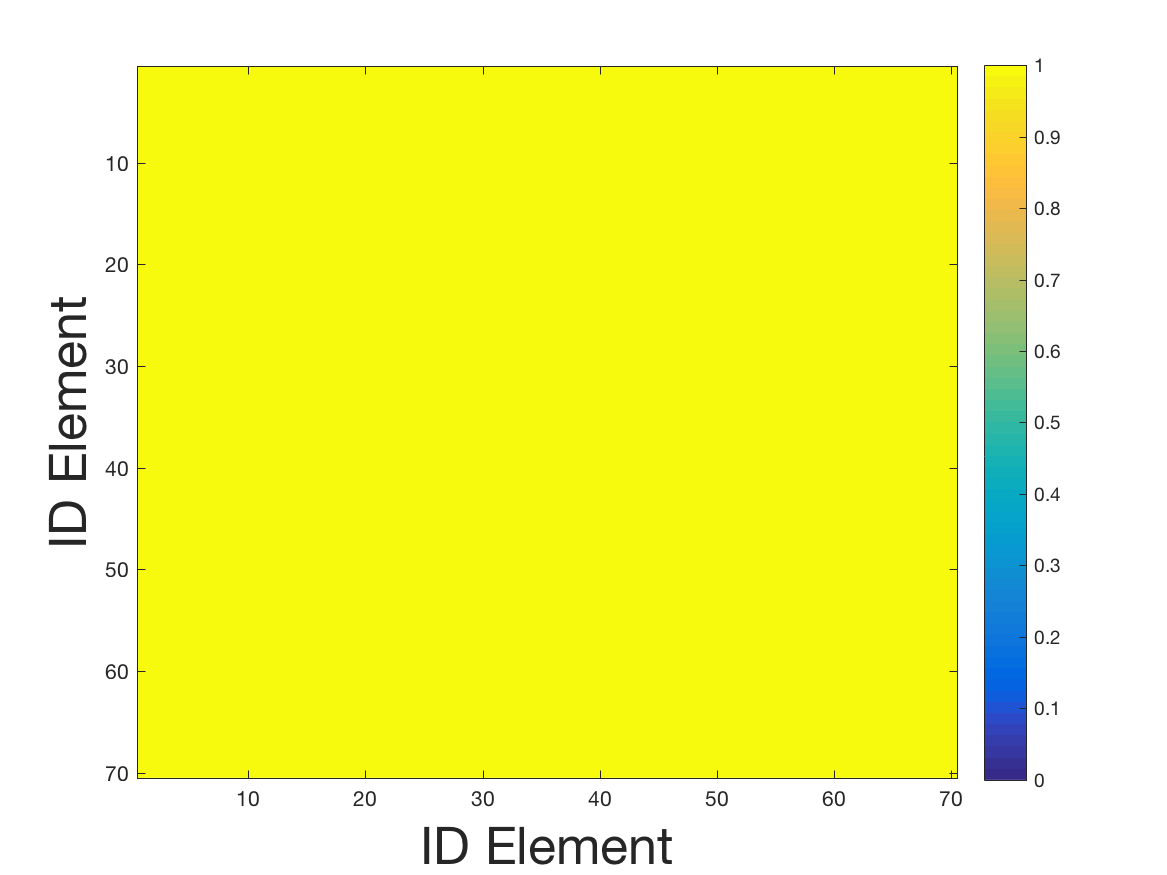}
    \includegraphics[width=0.45\textwidth]{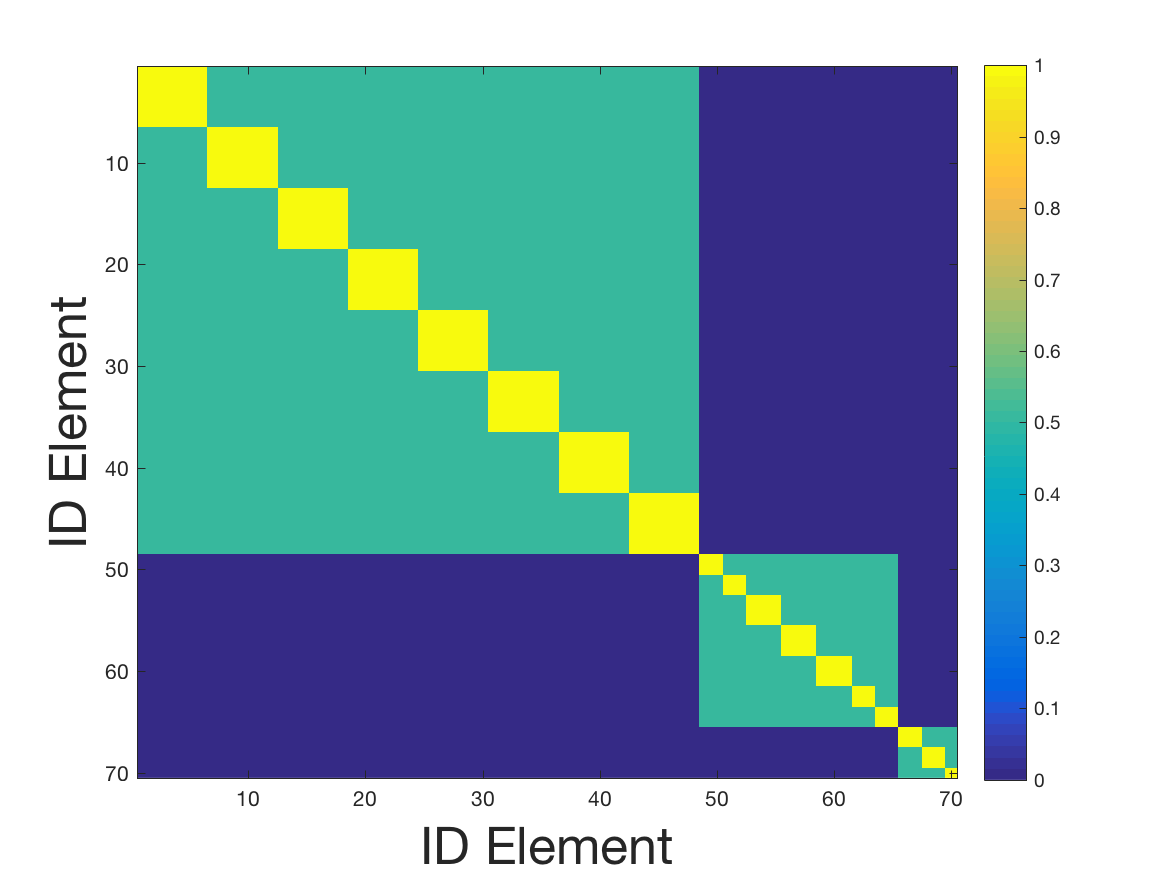}
    \includegraphics[width=0.45\textwidth]{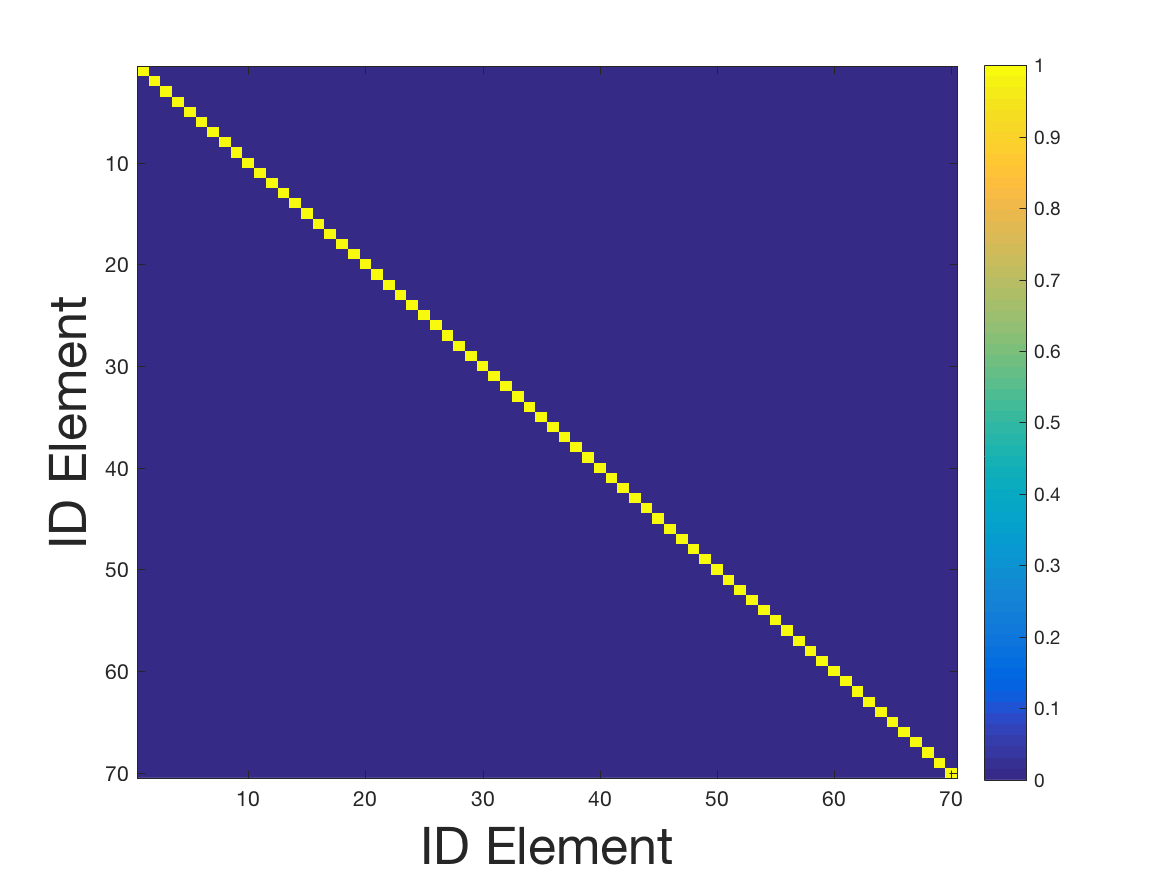}
    \caption{Correlation matrix $\rho$ among the 70 elements within the LFT, MFT and HFT that result from the classification in which, both, frequency and wafer, are considered. The four different cases considered as illustrative examples, Case 3.0, Case 3.1, Case 3.2, and Case 3.3, are plotted from top-left to bottom-right (see text for details).}
    \label{matrix_correl3}
\end{figure}

\begin{figure}[ht]
    \centering
    \includegraphics[width=1\textwidth]{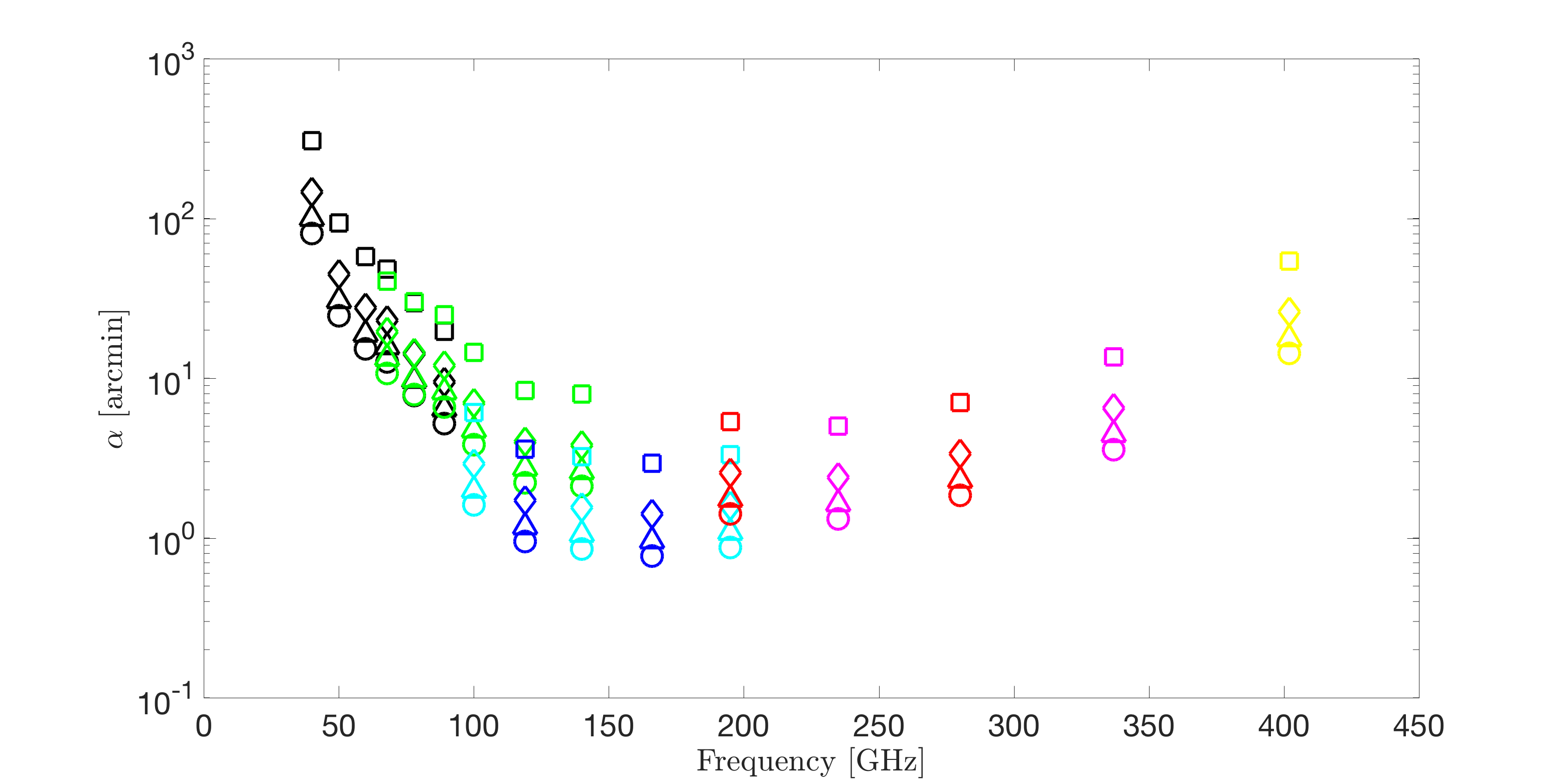}
    \caption{Polarization angle requirements for the 7 different types of wafers in the LFT, MFT and HFT telescopes. The four  correlation cases 3.0 (squares), 3.1 (circles), 3.2 (diamonds), and 3.3 (triangles) are plotted. The colours refer to the different types of wafers: black, type 1 of LFT; green, type 2 of LFT; dark-blue, type 1 of MFT, light-blue, type 2 of MFT; red, type 1 of HFT; pink, type 2 of HFT; yellow, type 3 of HFT.}
    \label{constraints_wafers_absolute}
\end{figure}

We represent the polarization angle requirements for the four cases and each one of the 7 different types of wafers in Fig.~\ref{constraints_wafers_absolute}.

\subsection{Relative angle requirements at detector level}
\label{sec:results_rel_det}

Let us finish this Section by discussing the requirements needed at detector level for LiteBIRD. In order to do that, we will follow the approach described in Subsection~\ref{sec:correlation_det}, starting from the constraints established at the frequency element level. In this case, we only need to use the number of detectors per frequency element, in order to infer a requirement on the polarization angle for each one of the detectors forming a given frequency element. This is given in the last column of Table~\ref{instrument}.

As an example, we will study the requirements for the three particular cases discussed in~\ref{sec:correlation_det}, and for which, we derived particular solutions of Eq.~\ref{eq:errors_equals}. These cases are:
\begin{itemize}
\item Case 4.0: The offsets of all the detectors and, therefore, all the frequency elements are correlated.
\item Case 4.1: The offsets of all the detectors and, therefore, all the frequency elements are uncorrelated.
\item Case 4.2: The offsets of all the detectors of a given frequency element are correlated, but frequency elements are uncorrelated among them.
\end{itemize}

Results, are summarized in Table~\ref{tab:alpha_relative_det}, where we present constraints for each detector of a given frequency element. For the most restrictive case (i.e., Case 4.0, full correlation), we obtain that the strongest constraints are reaching uncertainties of around 1 arcmin, for the most sensitive elements. This is similar to the frequency element constraints, as one would expect, since
we are imposing full correlation among detectors and frequency elements. The situation is much more relaxed for the case of full uncorrelation, having requirements of around 1 degree for the detectors at the most sensitive frequencies.
\begin{table}[]
    \centering
    \begin{tabular}{l|c|c|c|c}
    
         Element (Band) & \multicolumn{3}{c}{$\sigma_\alpha$ (arcmin)} \\
        (Telescope + Freq.) & Case 4.0 & Case 4.1 & Case 4.2 \\
        \noalign{\smallskip}
            \hline 
              \noalign{\smallskip}
LFT\_040GHz & 7.4 & 495.6 & 28.3  \\
LFT\_050GHz & 3.0 & 198.1 & 11.3  \\
LFT\_060GHz & 2.4 & 160.7 & 9.2  \\
LFT\_068GHz\_a & 6.3 & 420.9 & 24.0  \\	
LFT\_068GHz\_b & 2.7 & 178.6 & 10.2  \\
LFT\_078GHz\_a & 3.8 & 256.7 & 14.7  \\
LFT\_078GHz\_b & 1.9 & 129.5 & 7.4  \\
LFT\_089GHz\_a & 2.4 & 160.2 & 9.1  \\
LFT\_089GHz\_b & 2.2 & 146.5 & 8.4  \\
LFT\_100GHz & 1.7 & 113.5 & 6.5  \\
LFT\_119GHz & 0.9 & 62.8 & 3.6  \\
LFT\_140GHz & 0.8 & 55.8 & 3.2  \\
              \noalign{\smallskip}
MFT\_100GHz & 2.9 & 194.1 & 11.1  \\
MFT\_119GHz & 1.7 & 116.9 & 6.7  \\
MFT\_140GHz & 1.6 & 109.9 & 6.3  \\
MFT\_166GHz & 1.7 & 111.5 & 6.4  \\
MFT\_195GHz & 2.0 & 134.2 & 7.7  \\
              \noalign{\smallskip}
HFT\_195GHz & 3.1 & 206.5 & 11.8  \\
HFT\_235GHz & 3.3 & 218.1 & 12.5  \\
HFT\_280GHz & 5.3 & 356.7 & 20.4  \\
HFT\_337GHz & 13.4 & 902.3 & 51.5  \\
HFT\_402GHz & 83.6 & 5611.2 & 320.3	
    \end{tabular}
    \caption{Polarization angle requirements for each detector at each one of the 22 frequency elements within the LFT, MFT and HFT, for correlation cases 4.0, 4.1, and 4.2 (see text for details).}
    \label{tab:alpha_relative_det}
\end{table}

All the previous requirements, for absolute and relative angles and for the different schemes of grouping the frequency detectors, have been obtained assuming that the residual foregrounds are those estimated in~\cite{Errard2016}. As discussed at the end of Section~\ref{sec:method}, considering the contribution of residuals will always tend to decrease the polarization angle requirements. Its impact comes through the $A$ factor defined in Eq.~\ref{eq:A_factor}, and can be easily estimated resulting in the following changes in the angle requirements for all the cases considered: $18\%$ larger for negligible foreground residuals, and $10\%$ smaller for a level of residuals 3 times larger than the typical value estimated in~\cite{Errard2016} (see Fig.~\ref{residuals}).

\section{Conclusions and discussion}
\label{sec:conclusions}
We have presented a new methodology to establish requirements on the polarization angle accuracy that a set of  
detectors at a given frequency of a CMB experiment must satisfy, in order to ensure that the bias on the tensor-to-scalar ratio parameter $r$ is kept below a certain value. Note that the proposed work is focused on the estimation of requirements, while the establishment of a methodology to get them is out of the scope of this paper.

We consider different classifications of the detectors of a given frequency (not only on frequency elements), to explore in detail the implications at several levels: experiment as a whole, instruments, frequency elements, wafers, and individual detectors.

In particular, the method assumes that the CMB solution can be obtained through a linear combination of the different sets of detectors that observed the microwave sky at different frequency elements. Examples of this kind of CMB solutions are internal linear combination approaches (e.g, \texttt{PILC}), internal template fitting (e.g., \texttt{SEVEM}), or some kind of parametric fitting methods (e.g., \texttt{FGBuster}), used as a baseline for LiteBIRD\textcolor{blue}, which is the application  experiment shown in this work. An specific choice of the weights is particularly relevant and useful. In the reported case those coefficients are inversely proportional to the noise variance, providing a similar scheme to the one obtained with \texttt{FGBuster}. This work assumes this weighting scheme to provide final requirements on different detector sets for LiteBIRD, but the developed methodology is general.

Assuming that the requirements on the polarization angle that must satisfy the different sets of detectors are small enough to work on the small angle limit (a few degrees at most), we are able to obtain analytical solutions relating the bias on the $r$ parameter with the polarization angle uncertainties, and as a function of possible uncertainty correlations among the different detector sets. This analytical formalism provides absolute and relative polarization angle requirements.

As a practical application, we follow the criterion established in LiteBIRD, in such a way that the final uncertainty on the polarization angles do not induce biases on $r$ ($\delta_r$) larger than $5.77\times10^{-6}$, i.e., $1\%$ of the total uncertainty budget assigned to the whole systematic effects. With this constraint, we show that, at global and telescope levels, requirements of a few arcmin are needed, for the most restrictive case, in which all the angles are fully correlated. Requirements are relaxed a factor of 2 for the opposite case of angles fully uncorrelated.

At the frequency element level, the requirement on $\delta_r$ translate on requirements on the polarization angles a bit below 1 arcmin for the most sensitive frequencies (around 150 GHz) and a few tens of arcmin at the lowest and highest frequencies (around 40 and 400 GHz, respectively). Again, this is for the most restrictive case of full correlation among the polarization angles, obtaining constraints around 50\% more relaxed for the opposite case of non-correlated systematics.

We also consider the case in which the set of detectors are divided attending to the particular wafer division. In this case, we have 70 elements, accounting for detectors at different wafer type, frequency band and telescope. In particular, we can identify 7 different types of wafers. Again, requirements slightly below 1 arcmin are needed for the most sensitive frequencies, and ten times larger at the lowest and highest frequencies, for the most conservative case of full correlation. The requirements are relaxed a factor of 3 for the opposite case of non-correlations.

Additionally, from the requirements established at the frequency element level, we are also able to derive the constraints at the detector level, for each detector of the different frequency elements. In particular, for the case of full correlation we obtain similar requirements as those obtained from the frequency element level analyses, as one would expect. However, for the case of complete uncorrelation, we obtain more relaxed values for the requirements of several tens of arcmin. 

Finally, let us comment that these requirements appear to be achievable, as the first attempts made in \cite{IF_Pol} and \cite{delaHoz2021} seem to indicate. However, more complete analyses, including different correlation schemes among detectors or considering a detailed modellisation of the expected level of correlation for a given design of the instrument would be needed to fully demonstrate it.

\acknowledgments
This work is supported in \textbf{Japan} by ISAS/JAXA for Pre-Phase A2 studies, by the acceleration program of JAXA research and development directorate, by the World Premier International Research Center Initiative (WPI) of MEXT, by the JSPS Core-to-Core Program of A. Advanced Research Networks, and by JSPS KAKENHI Grant Numbers JP15H05891, JP17H01115, and JP17H01125. The \textbf{Italian} LiteBIRD phase A contribution is supported by the Italian Space Agency (ASI Grants No. 2020-9-HH.0 and 2016-24-H.1-2018), the National Institute for Nuclear Physics (INFN) and the National Institute for Astrophysics (INAF). The \textbf{French} LiteBIRD phase A contribution is supported by the Centre National d’Etudes Spatiale (CNES), by the Centre National de la Recherche Scientifique (CNRS), and by the Commissariat à l’Energie Atomique (CEA). The \textbf{Canadian} contribution is supported by the Canadian Space Agency. The \textbf{US} contribution is supported by NASA grant no. 80NSSC18K0132. 
\textbf{Norwegian} participation in LiteBIRD is supported by the Research Council of Norway (Grant No. 263011). The Spanish LiteBIRD phase A contribution is supported by the \textbf{Spanish} Agencia Estatal de Investigación (AEI), project refs. ESP2017-83921-C2-1-R, PID2019-110610RB-C21,  PID2020-120514GB-I00, ProID2020010108,and funding from Unidad de Excelencia María de Maeztu (MDM-2017-0765) co-funded with EU FEDER funds. Funds that support the \textbf{Swedish} contributions come from the Swedish National Space Agency (SNSA/Rymdstyrelsen) and the Swedish Research Council (Reg. no. 2019-03959). The \textbf{German} participation in LiteBIRD is supported in part by the Excellence Cluster ORIGINS, which is funded by the Deutsche Forschungsgemeinschaft (DFG, German Research Foundation) under Germany’s Excellence Strategy (Grant No. EXC-2094 - 390783311). 
We make use of the \texttt{CAMB}  \cite{Lewis2011} package.

\bibliographystyle{JHEP}
\bibliography{ref}

\appendix
\section{Gaussian Likelihood for $\delta_r$}
\label{sec:Appendix_A}
We discuss here the likelihood that allows one to obtain the error induced in the tensor-to-scalar parameter $r$ (Eq.~\ref{eq:delta_r}) by the rotation angles in Eq.~\ref{eq:CMB_power}.

Let us start by defining the bias in the B-mode CMB angular power spectrum, $\Delta B_{\ell}$, due to the presence of polarization angle offsets. This bias is given by subtracting from the final estimated B-mode CMB angular power spectrum (Eq.~\ref{eq:BB_obs})
the known contributions, without being aware of the polarization angle offsets: $r$ times the B-mode CMB angular power spectrum due to inflationary gravitational waves, $B_{\ell}^\mathrm{fid}$; the induced gravitational lensing from the E-modes, $L_{\ell}$; the residual foregrounds, $R_{\ell}^B$; and the effective noise, $N_{\ell}^\mathrm{eff}$, angular power spectra (Eq.~\ref{eq:noise_eff}). This subtraction defines Eq.~\ref{eq:BB_bias} that, for clarity, is repeated here:
\begin{equation}
    \label{eq:BB_bias_App}
    \Delta B_{\ell}=\left(rB_{\ell}^\mathrm{fid}+L_{\ell}+R_{\ell}^B\right) (\Sigma_{cos} -1 ) + (E_{\ell}+R_{\ell}^E)\Sigma_{sin} \ . 
\end{equation}

Therefore, this bias $\Delta B_{\ell}$ could introduce a bias on $r$ that can be characterized by the following Gaussian likelihood:
\begin{equation}
    \label{eq:likelihood}
    \log \mathcal{L} \propto -\frac{1}{2} 
     \sum_{\ell =2}^{\ell_{\mathrm{max}}}
     \frac{\left( \Delta B_{\ell} - \delta r B_{\ell}^\mathrm{fid} \right)^2}{\mathrm{Var}(B_{\ell})},
\end{equation}
where $\mathrm{Var}(B_{\ell})$ is the cosmic variance of the observed B-mode angular power spectrum, given by Eq.~\ref{eq:cosmic_var}. Solving this likelihood for $\delta r$, one obtains the analytical expression of Eq.~\ref{eq:delta_r}.

\section{Dealing with negative weights to build the CMB map}
\label{sec:Appendix_B}
During the discussion given in Section~\ref{sec:method_c}, in particular from Eq.~\ref{eq:delta_r_assumption}, it has been assumed that the weights of the linear combination of the frequency elements to derive the CMB signal were all positive. The aim of this Appendix is to provide a general formalism for weights $w$ that can take positive and negative values, as it is the case of  general ILC or template fitting methods~\cite{Fernandez-Cobos2016}.

Under the assumption that $\sigma_\nu=c|w_{\nu}|^{-1}$, Eq.~\ref{eq:delta_r_assumption} becomes:
\begin{equation}
    \label{eq:delta_r_no-assump}
     \langle \delta r \rangle = 4c^2\left[\sum_{\ell =2}^{\ell _{\mathrm{max}}}{\frac{(E_{\ell}+R_{\ell}^E)B_{\ell}^\mathrm{fid}}{\mathrm{Var}(B_{\ell})}}\right]
    \left[ \sum_{\ell =2}^{\ell _{\mathrm{max}}}
    {\frac{(B_{\ell}^\mathrm{fid})^2}{\mathrm{Var}(B_{\ell})}}
    \right]^{-1} \nonumber \\
    \Bigg[ 
        \sum_{\nu_1,\nu_2=1}^n  {\rho_{\nu_1 \nu_2} \frac{w_{\nu_1}w_{\nu_2}}{|w_{\nu_1}||w_{\nu_2}|}}\Bigg]\ \ .
\end{equation}

This is the general form for the expected bias in $r$, $\langle \delta r \rangle$. It is easy to show that, in the case in which all the $n$ polarization angles, $\alpha_\nu$, are fully correlated, i.e., $\rho_{\nu_1\nu_2}=1$ for any pair of frequency elements ($\nu_1$,$\nu_2$), Eq.~\ref{eq:delta_r_corr} takes now the form:
\begin{equation}
    \label{eq:delta_r_no-assump_corr}
     \langle \delta r \rangle = 4c^2\left[\sum_{\ell =2}^{\ell _{\mathrm{max}}}{\frac{(E_{\ell}+R_{\ell}^E)B_{\ell}^\mathrm{fid}}{\mathrm{Var}(B_{\ell})}}\right]
    \left[ \sum_{\ell =2}^{\ell _{\mathrm{max}}}
    {\frac{(B_{\ell}^\mathrm{fid})^2}{\mathrm{Var}(B_{\ell})}}
    \right]^{-1} \nonumber \\
    \Bigg[ 
        \sum_{\nu_1,\nu_2=1}^n  { \frac{w_{\nu_1}w_{\nu_2}}{|w_{\nu_1}||w_{\nu_2}|}}\Bigg]\ \ .
\end{equation}
For the case in which all the $n$ angles are uncorrelated, 
i.e. $\rho_{\nu_1\nu_2}=0$ for $\nu_1\neq \nu_2$, Eq.~\ref{eq:delta_r_random} is preserved.

\end{document}